\documentclass[superscriptaddress,preprintnumbers,amssymb,nofootinbib]{revtex4-2}

\usepackage{tikz-feynman}
\usepackage{subfig}
\usepackage{bm}
\usepackage{esvect}
\usepackage{verbatim}
\usepackage{multirow}

\usepackage[left]{lineno}
\usepackage{blindtext}
\pdfoutput=1
\usepackage[T1]{fontenc}
\usepackage{graphicx}
\usepackage{epsfig}
\usepackage{amsmath}
\usepackage{amsfonts}
\usepackage{amssymb}
\usepackage{braket}
\usepackage{cancel}
\usepackage{pstricks}
\usepackage{color}
\usepackage{feynmf}
\usepackage[normalem]{ulem}
\usepackage{epstopdf}
\usepackage{soul}
\usepackage{cancel}

\definecolor{ao(english)}{rgb}{0.0, 0.5, 0.0}

\usepackage{braket}
\usepackage{cancel}
\usepackage{setspace}
\usepackage{pstricks}
\usepackage{xcolor}
\usepackage{float}
\usepackage{mathtools}
\usepackage{multirow}
\usepackage{booktabs}
\usepackage{enumitem}
\providecommand{\tabularnewline}{\\}

\def\missET {{\not\!\! E_T}}
\def\noeft{{\color{gray}{{\footnotesize\sf [EFT?]}}}}
\def\gsim{\lower0.5ex\hbox{$\:\buildrel >\over\sim\:$}}
\def\lsim{\lower0.5ex\hbox{$\:\buildrel <\over\sim\:$}}

\begin{document}


\title{New flavor physics in di- and tri-lepton events from single-top at the LHC and beyond}

\author{Yoav Afik}
\email{yoavafik@campus.technion.ac.il}
\author{Shaouly Bar-Shalom}
\email{shaouly@physics.technion.ac.il}
\affiliation{Physics Department, Technion--Institute of Technology, Haifa 3200003, Israel}
\author{Amarjit Soni}
\email{adlersoni@gmail.com}
\affiliation{Physics Department, Brookhaven National Laboratory, Upton, NY 11973, US}
\author{Jose Wudka}
\email{jose.wudka@ucr.edu}
\affiliation{Physics Department, University of California, Riverside, CA 92521, USA}

\date{\today}

\begin{abstract}
The associated production of a single-top with opposite-sign same-flavor (OSSF) di-leptons, $pp \to t \ell^+ \ell^-$ and $ pp \to t \ell^+ \ell^- + j$ ($j=$light jet),
can lead to striking tri-lepton $pp \to \ell^\prime \ell^+ \ell^- + X$ and di-lepton $pp \to \ell^+ \ell^- + j_b + X$ ($j_b=b$-jet) events at the LHC, after the top decays.
Although these rather generic multi-lepton signals are flavor-blind, 
they can be generated by new 4-Fermi flavor changing (FC) $u_i t \ell \ell$ scalar, vector and tensor interactions ($u_i \in u,c$), which we study in this paper; we match the FC $u_i t \ell \ell$ 4-Fermi terms to the SMEFT operators and also to different types of FC underlying heavy physics.  The main backgrounds to these di- and tri-lepton signals arise from 
$t \bar t$, $Z$+jets and $VV$ ($V=W,Z$) production, but they can be essentially eliminated with a sufficiently high invariant mass selection on the OSSF di-leptons, $m_{\ell^+ \ell^-}^{\tt min}(OSSF) \gsim 1$ TeV; the use of $b$-tagging as an additional selection in the di-lepton final state case also proves very useful. We find, for example, that the expected 95\% CL bounds on the scale of a tensor(vector) $u t \mu \mu$ interaction, with the current $\sim 140$ fb$^{-1}$ of LHC 
data, are $\Lambda \lsim 5(3.2) $~TeV or $\Lambda \lsim 4.1(2.7)$~TeV, if analyzed via the di-muon $\mu^+ \mu^- + j_b$ signal or the $e \mu^+ \mu^-$ tri-lepton one, respectively.   
The expected reach at the HL-LHC with 3000 fb$^{-1}$ of data is $\Lambda \lsim 7.1(4.7)$~TeV 
and $\Lambda \lsim 2.4(1.5)$~TeV for the corresponding 
$u t \mu \mu$ and $c t \mu \mu$ operators. This should be compared to the current bounds of $\Lambda \lsim {\cal O}(1)$ TeV on both the $ut \ell \ell$ and $ct\ell \ell$ operators from LEP2 and from $pp \to t \bar t$ followed by $t \to \ell^+ \ell^- j$. We also study 
the potential sensitivity at future 27~TeV and 100~TeV high-energy LHC successors, which, for the $ut \ell \ell$ operators, can reach $\Lambda \sim 10-40$~TeV.
We furthermore discuss the 
possible implications of this class of FC 4-Fermi effective interactions on lepton non-universality tests at the LHC.
\end{abstract}

\maketitle
\flushbottom

\newpage 
\tableofcontents

\section{Introduction \label{sec:intro}}

The origin of the observed flavor pattern in the fermion sector still remains one of the fundamental  unresolved questions in theoretical particle physics. In particular, tree-level Flavor-Changing Neutral Currents (FCNC) are absent in the Standard Model (SM), so that FCNC effects in the SM are, in many cases, vanishingly small since they can only arise at the loop level and are GIM suppressed; this is the case for
$t \to u$ and $t \to c$ transitions in top decays~\cite{FCtopdecay1,FCtopdecay2,FCtopdecay3,FCtopdecay4,FCtopdecay5,FCtopdecay6,FCtopdecay7,FCtopdecay8,t_to_cdecay_soni,t_to_c_Hou} and/or top-production processes~\cite{FCtopprod1,FCtopprod2,FCtopprod3,FCtopprod4,FCtopprod5,eetc_soni1,eetc_soni2,eetc_Hou1}. 
Thus, the feeblest signal of FCNC effects in the top sector, either direct or indirect, may be an indicator of new flavor physics beyond the SM. This fact has led to a lot  of theoretical as well as experimental activity in understanding and searching for top FCNC within model independent approaches, as well as within specific popular models beyond the SM.

The significantly larger mass of the top-quark compared to all other quarks, best manifests the SM flavor problem and makes it the most sensitive to several types of New Physics (NP) and, in particular, to new flavor and CP-violation physics~\cite{ourreview}. For example, FCNC effects in decays of a quark will be typically suppressed by some power of $m_q/\Lambda$, where $\Lambda$ is the scale of the underlying NP, so that the larger the quark mass, the more significant the FCNC effects.
For this reason, searching for new FC dynamics in the top-sector was and is one of the major goals of past, current and future colliders. However, unfortunately, after more than a decade of collecting data and searching for NP in numerous processes at the 7, 8 and 13~TeV LHC, it is now clear that the scale of any natural underlying heavy physics and, in particular, the scale of possible flavor violation in the 3rd generation fermion sector, lies above $\Lambda \sim 1-2$~TeV. Indeed, even for decays of the top-quark, where the expected suppression factor for the corresponding NP-generated FC partial width is $(m_t/\Lambda)^n$ (typically $n = 4$ for FCNC top decays), the search for new underlying FC physics is extremely 
difficult.

On the other hand, the corresponding suppression factor for the cross-sections of any NP-generated FC scattering processes involving the top-quark, will be typically proportional to some power of $v/\Lambda$ or $E_{\tt cm}/\Lambda$, where $E_{\tt cm}$ is the c.m. energy of the collider. In particular, as we show below, the FC $t \to u$ and $t \to c$ transitions can be very efficiently studied in scattering processes, in some selected single-top production processes, where the FCNC effects are enhanced and SM backgrounds dramatically suppressed, 
i.e., at high $E_{\tt cm}$, which is particularly useful from the experimental point of view. 

Having emphasized the advantages of using scattering processes at the LHC as a testing ground for NP and, in particular, for searches of FC effects in top-quark systems, we now turn to a concrete illustration of these general statements. We will consider 
the following di-lepton and tri-lepton signals 
with a pair of opposite-sign same-flavor (OSSF) leptons:
\begin{eqnarray}
&& pp \to \ell^\prime \ell^+ \ell^- + X  ~,
\label{eq:lll_inclusive_intro} \\
&& pp \to \ell^+ \ell^- + j_b + X  ~,
\label{eq:lljb_intro}
\end{eqnarray}
where a selection of a single $b$-tagged jet is used with the di-lepton final state and, in general, $\ell,\ell^\prime = e,\mu$ or $\tau$ and $\ell^\prime = \ell$ and/or $\ell^\prime \ne \ell$ can be considered in the tri-lepton case.
These di- and tri-lepton signals are useful 
for generic NP searches and, as it turns out, although they are flavor-blind, they can also be very effectively used to search for FCNC physics in the top sector.

We study here   
the effects of higher dimensional effective 4-Fermi $t u_i \ell \ell$ FC interactions,\footnote{Some of the $t \to u_i$ 4-Fermi operators that we consider below are especially interesting, since by gauge invariance (see further discussion below), they also contribute to $b\to s \ell^+ \ell^-$ and $b\to c \ell^- \nu_{\ell}$ transitions and, therefore, to the anomalies observed in the ratios $R_{K^{(*)}}$ and $R_{D^{(*)}}$ in neutral and charged semileptonic  B-decays~\cite{Aaij:2014pli,Aaij:2014ora,Aaij:2017vbb,Aaij:2015esa,Aaij:2015oid,Wehle:2016yoi,Abdesselam:2016llu,ATLAS:2017dlm,CMS:2017ivg,Bifani:2017gyn,Aaij:2019wad,Abdesselam:2019wac,Lees:2012xj,Lees:2013uzd,Huschle:2015rga,Hirose:2016wfn,Aaij:2015yra,Aaij:2017uff,Aaij:2017deq,Adamczyk:2019wyt,Abdesselam:2019dgh} (for a recent review see~\cite{Bifani:2018zmi}). If confirmed, these anomalies would favor a multi-TeV scale for lepton-flavor non-universal (LFNU) new physics not only in $B$ decays, but also in the $ t \to u,c$ transitions studied in this paper.} where $u_i$ stands for either a $u$ or a $c$-quark and $\ell$ can be either of the three SM charged leptons, $\ell=e,\mu,\tau$.\footnote{We note that final states involving the $\tau$ have, in general, a lower experimental detection efficiency and are, therefore, expected to be less effective for our study.}
Specifically, we will show that the higher dimensional FC $t u_i \ell \ell$ operators are best   
studied via the following single-top + di-lepton associated production channels with 0 and/or 1 accompanying light-jet $j$ ($t$ stands for either a top or anti-top quark):\footnote{An interesting example of single-top production that can potentially lead to the di-lepton and tri-lepton signals in \eqref{eq:lll_inclusive_intro} and \eqref{eq:lljb_intro} was recently studied in \cite{tZprime}. They investigated the effects of FC $Z^\prime_\mu t_R \gamma^\mu u_R$ and $Z^\prime_\mu t_R \gamma^\mu c_R$ couplings on the process $pp \to t Z^\prime$, 
which can lead to the $(t \ell \ell)_0$ signal in \eqref{eq:llt} if the $Z^\prime$ also couples to a pair of SM leptons.}$^{,}$\footnote{We note that a large charge asymmetry is expected, e.g., in $pp \to \ell^{\prime +} \ell^+ \ell^-$ versus $pp \to \ell^{\prime -} \ell^+ \ell^-$, due to an asymmetric production of top versus anti-top quarks in \eqref{eq:llt} via 
$ug$-fusion (see Fig.~\ref{fig:Feynman}), which is caused by the asymmetric $u$ versus anti-$u$ quark densities in the LHC $pp$ initial state.} 
\begin{eqnarray}
(t \ell \ell)_0 &:& pp \to \ell^+ \ell^- + t ~,  \nonumber \\
(t \ell \ell)_1 &:& pp \to \ell^+ \ell^- + t  + j ~,
\label{eq:llt}
\end{eqnarray}
that lead to the di-lepton and tri-lepton signals in \eqref{eq:lll_inclusive_intro} and \eqref{eq:lljb_intro}, after the top decays via $t \to b W$ and, in the tri-lepton case, followed by $W \to \ell^\prime \nu_{\ell^\prime}$. Also, only the case of lepton flavor diagonal $tu_i \ell \ell$ 4-Fermi contact terms will be studied, so that the di-leptons $\ell \ell$ in \eqref{eq:llt} and therefore also in \eqref{eq:lll_inclusive_intro} and \eqref{eq:lljb_intro} are OSSF.
Note, though, that similar effects are expected from the lepton flavor violating $tu_i \ell \ell^\prime$ 4-Fermi interactions if the underlying scale of lepton flavor violation is also at the multi-TeV scale, see e.g.,~\cite{topdecay3,topdecay2,Gottardo:2019lmv}.
Indeed, the presence of two-three high-$p_T$ charged leptons allows to have an efficient trigger 
strategy on such final 
states that can be used to very effectively cut down the event rate of the background. 
As will be shown, the new FC 4-Fermi $tu_i \ell \ell$ interactions can be isolated from the SM background, as well as from other potential sources of NP that can affect these $t \ell \ell$ signals, by looking at the off-Z peak behaviour of the OSSF di-leptons in the $(t \ell \ell)_0 \to \ell^+ \ell^- + j_b, ~ \ell^\prime \ell^+ \ell^-$ and $(t \ell \ell)_1 \to \ell^+ \ell^- + j_b, ~ \ell^\prime \ell^+ \ell^-$ signals from \eqref{eq:llt}.

\begin{figure}[htb]
  \centering
\includegraphics[width=0.70\textwidth]{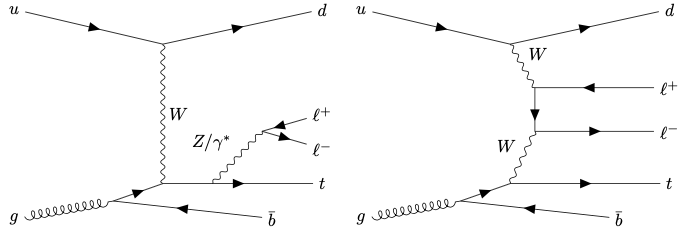}
\caption{Representative lowest-order Feynman diagrams for the SM single top-quark + di-lepton production 
with one light jet, $pp \to t \ell^+ \ell^- j$. Diagrams are shown for the
on-Z peak (left) and non-resonant $\ell^+ \ell^-$ (right) production cases.}
\label{fig:SM}
\end{figure}

Let us recall that, in the SM, single-top production at the LHC proceeds via several channels with 
different underlying leading topologies:
%
\begin{itemize}
    \item The so-called s-channel and t-channel $W$-exchange processes: $u \bar d \to t \bar b$ and $qb \to q^\prime t$ ($q,q^\prime=u,d$), respectively.
    \item Single top production in association with a gauge-boson, which are initiated by b-quarks in the proton: $bg \to t W$ and $bW \to tZ/\gamma$. In the four flavor scheme, where only light quarks and gluons are allowed in the initial state,  these processes are responsible for $tV+$jets 
    production, e.g., $pp \to tZ j+j_b$, where $j_b$ stands for a $b$-jet (see Fig.~\ref{fig:SM}). 
    \item Single top-Higgs associated production: $b W \to th$, which in the four flavor setup yields $pp \to thj$ and $pp \to thW$. 
\end{itemize}

All the single-top channels mentioned above with the exception of the $tW$ and $tWj$ final states are pure electroweak (EW) processes. It is for this reason that these channels, including possible EFT FC effects, have been widely studied in the past two decades~\cite{Degrande:2018fog,Maltoni-global,Hartland:2019bjb,Durieux:2019rbz,singletop1,tZgamma1,SMEFTtop1,tZ3,SMEFTtop2,SMEFTtop3,SMEFTtop4,Zhang:2016omx,tH1,tH2,tH3,tH4,tH5,tH6,tH7,tH8,tZ1,tZ2,tZ4,tZ5,tZ6,tZ7,tZ8-trilepton,tZ9-trilepton,tgamma1,tZEFT-decay1,tZprime}; a global and comprehensive analyses of the effects of various types of higher dimensional operators involving the top-quark field(s), including FC 4-Fermi operators of the type considered here (e.g., in~\cite{Maltoni-global}) can be found in~\cite{Maltoni-global,Hartland:2019bjb,Durieux:2019rbz,SMEFTtop1,tZ3,SMEFTtop2,SMEFTtop3,SMEFTtop4}. The effects of (2-quarks)(2-leptons) 4-Fermi operators (which are of interest in this study) on the single top + W production channel $pp \to t W \to \ell^+ \ell^- + j_b + \missET$, had been recently studied in~\cite{Tonero:2020zcy}, where bounds at the level of $\Lambda \gsim {\rm few} \times 100$~GeV were found on the scale of these operators.
We note also the study in~\cite{1008.3562} of the effects of FC 4-quarks operators in single-top production via $ q \bar q \to t + j$ ($j$=light-quark jet). 

Our $(t \ell \ell)_0$ zero-jets single-top + di-lepton channel has, therefore, no significant, irreducible SM tree-level contribution: the process requires a FC $t \to u$ insertion, and the leading order SM diagrams for this process are 1-loop and are GIM suppressed. The combination of these effects renders the corresponding amplitude unobservably small within the SM.  On the other hand, the $(t \ell \ell)_1$ channel $pp \to t \ell^+ \ell^- j$ does have potentially significant SM contributions~\cite{Campbell:2013yla,Pagani:2020mov,singletop-minirev,singletop-minirev-LHC,singletop-minirev-LHC2}, which is dominated by the EW associated production of a single-top with a $Z$-boson and an accompanying light-jet, i.e., via $u b \to t Z j$ in the five-flavor scheme, followed by the decay $Z \to \ell^+ \ell^-$ as shown in the left diagram of Fig.~\ref{fig:SM}. There is also a non-resonant contribution to $(t \ell \ell)_1$ (also depicted in Fig.~\ref{fig:SM}) which is, however, sub-leading in the SM, consisting of no more than $\sim 15 \%$ of the total cross-section~\cite{Pagani:2020mov}. The process $(t \ell \ell)_1$ has been measured by both ATLAS~\cite{Aaboud:2017ylb,Aad:2020wog} and CMS~\cite{Sirunyan:2017nbr,Sirunyan:2018zgs} collaborations, who focused on the on-$Z$ peak events, using a selection of $|m_{\ell^+ \ell^-} - m_Z| < 10$~GeV for the signal region. The total (full phase-space) cross-section was obtained by an extrapolation using the efficiency and acceptance factors calculated for the SM kinematics. In a very recent search by CMS \cite{Sirunyan:2020tqm}, the effects of 4-Fermi $t \bar t \ell^+ \ell^-$ operators (rather than the FC $t u_i \ell^+ \ell^-$ operators relevant to our study) on the $(t \ell \ell)_1$ process and other 
top(s) + di-lepton signals were studied (e.g., in $p p \to t \bar t \ell^+ \ell^-$), where off-Z peak di-lepton events were also considered with a selection $|m_{\ell^+ \ell^-} - m_Z| > 10$~GeV, though they did not make use of the "hard" selection $m_{\ell^+ \ell^-}(OSSF) >  1000$~GeV that we utilize in this work.

Indeed, our main interest in this paper will be the potential NP effects that contribute to the OSSF cross-section in the region of high di-lepton invariant masses, e.g., $m_{\ell^+ \ell^-}(OSSF) \gsim  (100-1500)$~GeV. This will be the case, in particular, for the EFT contributions we study below. As we show below, large deviations from the SM are expected also off the Z-peak in the $(t \ell \ell)_0$ and $(t \ell \ell)_1$ single-top + di-lepton channels of \eqref{eq:llt}, in the presence of new top-quark couplings to leptons, which do not necessarily involve anomalous couplings of the SM gauge-bosons to the top-quark. 

\section{New physics setup and Single-top + di-lepton production at the LHC} 

The NP will be parameterized by higher dimensional, gauge-invariant effective operators, ${\cal O}_i^{(n)}$, in the so-called SM Effective Field Theory (SMEFT) framework~\cite{EFT1,EFT2,EFT3,EFT4,EFT5}; the effective operators are constructed using the SM fields and their coefficients are suppressed by inverse powers of the NP scale $\Lambda$
\cite{EFT1,EFT2,EFT3,EFT4,EFT5}:
\begin{eqnarray}
{\cal L} = {\cal L}_{SM} + \sum_{n=5}^\infty
\frac{1}{\Lambda^{n-4}} \sum_i \alpha_i O_i^{(n)} \label{eq:EFT1}~,
\end{eqnarray}
where $n$ is the mass dimension of $O_i^{(n)}$ and we assume decoupling and weakly-coupled heavy NP, so that
$n$ equals the canonical dimension. The dominating NP effects are then expected to be generated by contributing operators with the lowest dimension (smallest $n$) that can be generated at tree-level in the underlying theory. The (Wilson) coefficients $\alpha_i$ depend on the details of the underlying heavy theory and, therefore, they parameterize all possible weakly-interacting and decoupling types of heavy physics; an example of matching this EFT setup to a specific underlying heavy NP scenario will be given below. 
 
\begin{figure}[H]
  \centering
\includegraphics[width=0.70\textwidth]{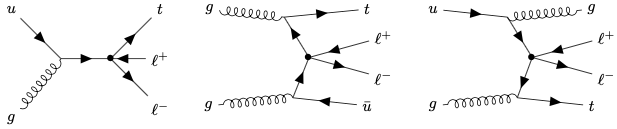}
\caption{Representative Feynman diagrams for the lowest order single top-quark + di-lepton production channels with no light jets $pp \to t \ell^+ \ell^-$ (left) and with one light jet $pp \to t \ell^+ \ell^- + j$ (middle and right) at the LHC, via the $t \bar{u} \ell^+ \ell^-$ 4-Fermi interaction (marked by a heavy dot).}
  \label{fig:Feynman}
\end{figure}

The dimension six operators ($n=6$) include seven 4-Fermi operators, listed in Table~\ref{tab:LNU-dim6}, that involve $t$ and $u$ quarks and a pair of charged leptons and are relevant for the processes we consider. As will be discussed below, these operators may also generate LFNU effects. In Fig.~\ref{fig:Feynman} we depict representative diagrams for the $(t\ell \ell)_{0}$ and $(t\ell \ell)_{1}$ processes in \eqref{eq:llt}, which are mediated by the $t \bar u \ell^+ \ell^-$ 4-Fermi operators in Table~\ref{tab:LNU-dim6}.

\begin{table*}[htb]
\caption{\label{tab:LNU-dim6}
The dimension six operators in the SMEFT, which potentially involve 
FC ($t \to u$) interactions between top-quarks and leptons and may, therefore, be
a source for lepton non-universal effects (see also text). 
The subscripts $p,r,s,t$ are flavor indices.}
\begin{center}
\small
\begin{minipage}[t]{4.45cm}
\renewcommand{\arraystretch}{1.5}
\begin{tabular}[t]{c|c}
\multicolumn{2}{c}{$4-{\rm Fermi}:(\bar LL)(\bar LL)$} \\
\hline
${\cal O}_{lq}^{(1)}(prst)$                & $(\bar l_p \gamma_\mu l_r)(\bar q_s \gamma^\mu q_t)$ \\
${\cal O}_{lq}^{(3)}(prst)$                & $(\bar l_p \gamma_\mu \tau^I l_r)(\bar q_s \gamma^\mu \tau^I q_t)$
\end{tabular}
\end{minipage}
\begin{minipage}[t]{4.45cm}
\renewcommand{\arraystretch}{1.5}
\begin{tabular}[t]{c|c}
\multicolumn{2}{c}{$4-{\rm Fermi}:(\bar RR)(\bar RR)$} \\
\hline
${\cal O}_{eu}(prst)$                      & $(\bar e_p \gamma_\mu e_r)(\bar u_s \gamma^\mu u_t)$ \\
\end{tabular}
\end{minipage}
\begin{minipage}[t]{4.45cm}
\renewcommand{\arraystretch}{1.5}
\begin{tabular}[t]{c|c}
\multicolumn{2}{c}{$4-{\rm Fermi}:(\bar LL)(\bar RR)$} \\
\hline
${\cal O}_{lu}(prst)$               & $(\bar l_p \gamma_\mu l_r)(\bar u_s \gamma^\mu u_t)$ \\
${\cal O}_{qe}(prst)$               & $(\bar e_p \gamma^\mu e_r)(\bar q_s \gamma_\mu q_t)$ \\
\end{tabular}
\end{minipage}

\vspace{0.25cm}

\begin{minipage}[t]{4.45cm}
\renewcommand{\arraystretch}{1.5}
\begin{tabular}[t]{c|c}
\multicolumn{2}{c}{$4-{\rm Fermi}:(\bar LR)(\bar L R)+\hbox{h.c.}$} \\
\hline
${\cal O}_{lequ}^{(1)}(prst)$ & $(\bar l_p^j e_r) \epsilon_{jk} (\bar q_s^k u_t)$ \\
${\cal O}_{lequ}^{(3)}(prst)$ & $(\bar l_p^j \sigma_{\mu\nu} e_r) \epsilon_{jk} (\bar q_s^k \sigma^{\mu\nu} u_t)$
\end{tabular}
\end{minipage}
\end{center}
\end{table*}
%
We will henceforward adopt the parameterization used in~\cite{our_tc_paper} for the effective Lagrangian of the FC $t \bar u \ell^+ \ell^- $ contact interactions (a similar parameterization for the $ t \bar t \ell^+ \ell^-$ interactions has been used in~\cite{Grzadkowski:1995te,Grzadkowski:1997cj} for the study of $e^+ e^- \to t \bar t$), which was also used by the DELPHI~\cite{DELPHI} and L3~\cite{L3} collaborations at LEP2 to set bounds on the $tcee$ contact interactions resulting from the 4-Fermi operators of Table~\ref{tab:LNU-dim6} (see also discussion below):
\begin{eqnarray}
{\cal L}_{tu\ell \ell} =  {1\over\Lambda^2} \sum_{i,j=L,R} \biggl[ V_{ij}^\ell \left({\bar \ell} \gamma_\mu P_i \ell \right) \left( \bar t \gamma^\mu P_j u \right)  + S_{ij}^\ell \left( {\bar \ell} P_i \ell \right) \left( \bar t P_j u \right)  + T_{ij}^\ell \left( {\bar \ell} \sigma_{\mu \nu} P_i \ell \right) \left( \bar t \sigma_{\mu \nu} P_j u \right) \biggr] \label{4fermimatrix}~,
\end{eqnarray}
where $P_{L,R} = (1 \mp \gamma_5)/2$ and $u$ represents a 1st or 2nd generation up-quark. In terms of the coefficients of the effective operators in Table~\ref{tab:LNU-dim6}, the vector-like ($V_{ij}^\ell$), scalar-like ($S_{ij}^\ell$) and tensor-like ($T_{ij}^\ell$) couplings are given by (we henceforward 
drop the superscript $\ell$):
\begin{eqnarray}
&& 
V_{LL}= \alpha_{\ell q}^{(1)} - 
\alpha_{\ell q}^{(3)} ~,~ 
V_{LR}= \alpha_{\ell u} ~, ~
V_{RR}= \alpha_{e u} ~, ~
V_{RL}= \alpha_{q e} 
~, \cr
&& 
S_{RR}= -\alpha_{\ell e q u}^{(1)} ~, ~
S_{LL}=S_{LR}=S_{RL}=0 
~, \cr
&& 
T_{RR}=-\alpha_{\ell e q u}^{(3)} ~, ~
T_{LL}=T_{LR}=T_{RL}=0 \label{vst}~.
\end{eqnarray}

These 4-Fermi interactions can be generated through tree-level exchanges of heavy vectors and scalars in the underlying heavy theory (or their Fierz transforms). Note that no $LL$ tensor or $LL$, $LR$ and $RL$ scalar terms are generated at dimension 6; they can, however, be generated by dimension 8 operators and thus have coefficients suppressed by $\sim ( v^2 /\Lambda^4)$, where $v=246$ GeV is the Higgs vacuum expectation value.

\subsection{Examples of matching to underlying beyond the SM  scenarios}
Interesting examples of underlying heavy particle tree-level exchanges that can generate some of the $tu_i \ell \ell$ operators above include  the $R_2$-type scalar Leptoquark (this is the only scalar Leptoquark that does not induce proton decay) and the $U_1$-type vector Leptoquark,  which transforms as $(3,2,7/6)$ and $(3,1,2/3)$ under the $SU(3) \times SU(2) \times U(1) $ SM gauge group, respectively.  These two Leptoquarks can address both $R_{K^{(*)}}$ and $R_{D^{(*)}}$ anomalies as well as the muon $g-2$ one (see~\cite{R2_1,R2_2,R2_3,R2_4,R2_5,R2_6,R2_7,R2_n} for the $R_2$ case and~\cite{U1-1,U1-2,U1-3,U1-4,U1-5,U1-6} for the $U_1$ case), having  the following couplings to a quark-lepton pair~\cite{Dorsner:2016wpm} :
\begin{eqnarray}
{\cal L}_{Y}^{R_2} &=& z {\bar e} R_2^{i \star} q^{i} - y {\bar u} R_2^i \epsilon_{ij} \ell^{j} + \text{h.c.} ~, \\
{\cal L}_{Y}^{U_1} &\supset& x {\bar q} \gamma_\mu U_1^\mu \ell + \text{h.c.} ~,
\end{eqnarray}
where $i,j$ are $SU(2)$ indices and flavor indices are not specified ($U_1$ can have additional $d_R \gamma_\mu e_R$ and $u_R \gamma_\mu \nu_R$ couplings which are not relevant to our setup). In particular, tree-level exchanges of $R_2$ and $U_1$ among the lepton-quark pairs induce (after a Fierz transformation)~\cite{R2_n,Gherardi:2019zil}:
\begin{eqnarray}
U_1&:& \alpha_{\ell q}^{(1)} = \alpha_{\ell q}^{(3)} = -\frac{x x^\star}{ 2 M_{U_1}^2} ~, \nonumber \\
R_2&:& \alpha_{qe} = -\frac{z z^\star}{ 2 M_{R_2}^2}~,~ \alpha_{\ell u} = \frac{y y^\star}{ 2 M_{R_2}^2}~,~ \alpha_{\ell e qu}^{(1)} = 4 \alpha_{\ell e qu}^{(3)} = -\frac{y z}{ 2 M_{R_2}^2} ~,
\end{eqnarray}
where $M_{R_2}$ and $M_{U_1}$ are the masses of $R_2$ and $U_1$, respectively.
Thus, following our parameterization in Eq.~\ref{4fermimatrix}, we see that the 4-Fermi vector couplings $V_{RL}$ and $V_{LR}$ as well as the scalar and tensor couplings $S_{RR}$ and $T_{RR}$ can be generated in the underlying heavy theory if it includes the Leptoquarks $R_2$, and if this Leptoquark couples e.g., to top-muon and up-muon (or charm-muon) 
pairs. 
It is interesting to note that, although $U_1$ contributes to the operators 
${\cal O}_{\ell q}^{(1)}$ and ${\cal O}_{\ell q}^{(3)}$, it doesn't generate the $V_{LL}$ vector interactions of \eqref{4fermimatrix}, since $\alpha_{\ell q}^{(1)} = \alpha_{\ell q}^{(3)}$ if ${\cal O}_{\ell q}^{(1)}$ and ${\cal O}_{\ell q}^{(3)}$ are generated by $U_1$. 
On the other hand, 
it will generate the $V_{LL}$ terms for the corresponding down-quark operators, e.g., $\left({\bar \ell} \gamma_\mu P_i \ell \right) \left( \bar b \gamma^\mu P_j s \right)$, for which 
$V_{LL} = \alpha_{\ell q}^{(1)} + \alpha_{\ell q}^{(3)}$, see e.g., \cite{Gherardi:2019zil}.

A compilation of the various types of NP that can induce the dimension six 4-Fermi interactions in Table~\ref{tab:LNU-dim6} can be found in~\cite{topdecay2}.

\section{Bounds and related phenomenology}

We now briefly summarize the current  bounds and phenomenology aspects related to the $t u_i \ell\ell$ 4-Fermi contact interactions of Eq.~\ref{4fermimatrix}. 

\subsection{The $tu_iee$ 4-Fermi operators involving two electrons}    
 
These operators can contribute to single top-quark + light-jet production at an $e^+ e^-$ machine: $e^+ e^- \to t + j$, where the light-jet $j$ originates from either a $u$ or a $c$-quark. Accordingly, these operators were studied and constrained at LEP2 by the DELPHI~\cite{DELPHI} and L3~\cite{L3} collaborations, who reported bounds ranging from $\Lambda \gsim 600$~GeV to $\Lambda \gsim 1.4$~TeV, depending on the underlying NP mechanism, i.e., whether a scalar, vector or tensor-like $tu_i ee$ 4-Fermi vertex is involved, and assuming  ${\cal O}(1)$ couplings for these interactions. A slight improvement can be obtained by combining these LEP2 bounds with (the rather weak) bounds derived from the rare top decay to a pair of charged leptons and a jet $t \to \ell^+ \ell^- j$~\cite{Maltoni-global,topdecay1,1008.3562,topdecay3,topdecay2}.\footnote{The partial FC top decay width $\Gamma_{\ell \ell u}=\Gamma(t \to \ell^+ \ell^- u)$ due to the 4-Fermi $t u_i \ell \ell$ scalar, vector and tensor interactions of \eqref{4fermimatrix} is: $\Gamma_{\ell \ell u} = (2 \pi m_t/3)  [m_t/(8 \pi \Lambda)]^4  \cdot \left( S_{RR}^2 + 4 \sum V_{ij}^2 +48 T_{RR}^2 \right)$~\cite{1008.3562,topdecay2}.}

\subsection{The $tu_i\mu\mu$ 4-Fermi operators involving two muons}  

The constraints on these operators are weak due to the absence of experimental bounds off the Z peak. In particular, bounds on these operators can be derived from $p p\to t \bar t$ production at the LHC, followed by $t \to \ell^+\ell^- j$ by one of the top-quarks, but no off-Z peak data was analysed in this channel. Note, however, the recent interesting analysis performed in~\cite{topdecay2} extending  existing $t \to Zj$ experimental searches in $t \bar t$ production at the LHC, using an off-Z peak dilepton invariant mass selection to put new bounds on the scale of $tu_i \ell \ell$ 4-Fermi operators of Table~\ref{tab:LNU-dim6}. They found e.g., that $\Lambda \gsim 0.8,1.0,1.5$~TeV can be reached at the future HL-LHC on the scalar, vector and tensor $tu \mu \mu$ and $tc \mu \mu$ interactions, respectively, for ${\cal O}(1)$ couplings: $S_{RR}=V_{ij}=T_{RR} =1$. These bounds are comparable to the LEP2 bounds on  $tu ee$ and $tce e$ discussed above, but, as we will show below, fall short by a factor of 3-5 compared to the sensitivity that can be obtained using the single-top + dilepton channels considered in this work.

\subsection{Implications of gauge invariance: consequences for $b$-quark scattering and $B$-physics} 

In operators involving left-handed quark isodoublets gauge invariance relates the 
$t u \ell \ell$ and  $b d \ell \ell$ 4-Fermi FC interactions.\footnote{Note that the correlation between operators involving the top-quark and operators involving the $b$-quark should be taken with caution, since sign differences can lead to e.g., a cancellation of effects for operators involving $b_L$ and an enhancement for those involving $t_L$ (or vice-versa).}  In particular, among the operators in Table~\ref{tab:LNU-dim6}, the $(\bar LL)(\bar LL)$ operators ${\cal O}_{l q}^{(1)}$, ${\cal O}_{l q}^{(3)}$ and the $(\bar LR)(\bar RL)$ one ${\cal O}_{qe}$,  include also the corresponding FCNC $b d \ell \ell$ interactions: 
\begin{eqnarray}
{\cal O}_{lq}^{(1)}(pr31) & = & (\bar\ell_p \gamma_\mu P_L \ell_r) \cdot \left[ (\bar t \gamma^\mu P_L u) + (\bar b \gamma^\mu P_L d) \right] ~, \nonumber \\
{\cal O}_{lq}^{(3)}(pr31) & \supset & 
(\bar\ell_p \gamma_\mu \tau^3 P_L \ell_r) \cdot \left[ (\bar t \gamma^\mu P_L u) - (\bar b \gamma^\mu P_L d) \right] ~, \nonumber \\
{\cal O}_{qe}(pr31)  & = & (\bar\ell_p \gamma^\mu P_R \ell_r) \cdot \left[ (\bar t \gamma_\mu P_L u)
+ (\bar b \gamma_\mu P_L d) \right] ~. 
\end{eqnarray}

Referring to \eqref{4fermimatrix} it then follows that the $V_{LL}$ and $V_{RL}$ couplings for the $t$ and $b$ quarks are related: $ V_{LL}(tu\ell\ell) = \alpha_{\ell q}^{(1)} - \alpha_{\ell q}^{(3)}$, $ V_{LL}(bd\ell\ell) = -\alpha_{\ell q}^{(1)} - \alpha_{\ell q}^{(3)}$ and $ V_{RL}(tu\ell\ell) =  V_{RL}(bd\ell\ell) = \alpha_{qe} $, and the corresponding scales $ \Lambda $ are the same. Similar relations occur for operators involving left-handed quarks of the 2nd and 3rd generations, e.g., $ V_{RL}(tc\ell\ell) =  V_{RL}(bs\ell\ell) $. 

The triplet operator ${\cal O}_{l q}^{(3)}(pr31)$ also includes the 4-Fermi charged currents, e.g., for the muon case: $(t \gamma_\mu P_L d)(\mu \gamma^\mu P_L \nu_\mu)$ and $(b \gamma_\mu P_L u)(\mu \gamma^\mu  P_L \nu_\mu)$, and, similarly,  ${\cal O}_{l q}^{(3)}(pr32) \supset  (t \gamma_\mu P_L s)(\mu \gamma^\mu P_L \nu_\mu), (b \gamma_\mu P_L c)(\mu \gamma^\mu P_L \nu_\mu)$. Furthermore, the $(\bar LR)(\bar LR)$ scalar and tensor operators ${\cal O}_{l equ}^{(1)}(pr31)$ and ${\cal O}_{l equ}^{(3)}(pr31)$ induce the charged currents involving the $b$-quark:  $(\bar b P_R u)(\bar\nu_\mu P_R \mu)$,  $(\bar b \sigma^{\mu \nu} P_R u)(\bar\nu_\mu \sigma_{\mu \nu} P_R \mu)$ and similarly the $b \to c$ ones for ${\cal O}_{l equ}^{(1)}(pr32)$ and ${\cal O}_{l equ}^{(3)}(pr32)$. 

Therefore, the $V_{LL}$, $V_{RL}$, $S_{RR}$ and $T_{RR}$ 4-Fermi $tu \ell \ell$ and $tc \ell \ell$ terms in 
\eqref{4fermimatrix}, have also interesting repercussions in scattering processes involving the b-quark in the final state, e.g., $dg \to b \ell \ell$~\cite{soniRPV,bsll,bbll,Altmannshofer:2020axr,our_LFU_paper}, in $b q$ scattering e.g., $b d \to \ell \ell, ~ bu \to \ell \nu_\ell$~\cite{Marzocca,Marzocca:2020ueu,Admir} as well as in B-decays (see e.g.~\cite{U1-2,Bordone:2018nbg}). For the latter, some of the notable ones include $B^+ \to \pi^+ \mu^+ \mu^-$, $B^+ \to K^+ \mu^+ \mu^-$ and 
$B^0_{d,s} \to \mu^+ \mu^-$ associated with $b \to d$ transitions,  see e.g., the recent analysis in~\cite{Gherardi:2019zil}, as well as   the $R_{K^{(*)}}$ and $R_{D^{(*)}}$ anomalies~\cite{Aaij:2014pli,Aaij:2014ora,Aaij:2017vbb,Aaij:2015esa,Aaij:2015oid,Wehle:2016yoi,Abdesselam:2016llu,ATLAS:2017dlm,CMS:2017ivg,Bifani:2017gyn,Aaij:2019wad,Abdesselam:2019wac,Lees:2012xj,Lees:2013uzd,Huschle:2015rga,Hirose:2016wfn,Aaij:2015yra,Aaij:2017uff,Aaij:2017deq,Adamczyk:2019wyt,Abdesselam:2019dgh,Bifani:2018zmi}, which occur in $b\to s \ell^+ \ell^-$ and $b\to c \ell^- \nu_{\ell}$ transitions, respectively, and may, therefore, be closely related to the $t u_i \ell \ell$ dynamics discussed in this work (see also~\cite{Kamenik:2018nxv,Camargo-Molina:2018cwu,London:2019ulu}). In particular, a best fit to $R_{K^{(*)}}$, $R_{K}$ and $B^0_{s} \to \mu^+ \mu^-$ observables implies that the scale of  
${\cal O}_{l q}^{(1)}(pr32)$ or ${\cal O}_{l q}^{(3)}(pr32)$ (or both) is around 
$40$~TeV~\cite{Gherardi:2019zil} assuming no cancellations, in which case the contribution of these operators to our single-top production processes is 
too small to be observed at the 13~TeV LHC. Alternatively, if single top production effects involving ${\cal O}_{l q}^{(1,3)}$ are observed at the LHC, this would indicate not only the presence of NP, but also that cancellations do in fact occur ($\alpha_{l q}^{(1)} \simeq \alpha_{l q}^{(3)} $), giving additional information about the properties of the new physics involved.

\subsection{Implications of gauge invariance: $ pp \to t +\missET$ and 
$ pp \to t \ell +\missET$ single-top signals}

The $(\bar LL)(\bar L L)$ vector operator ${\cal O}_{l q}^{(3)}$ as well as the $(\bar LR)(\bar L R)$ scalar and tensor 4-Fermi operators ${\cal O}_{lequ}^{(1)}$ and ${\cal O}_{lequ}^{(3)}$, which contribute to the FCNC $t u_i \ell \ell$ interactions ($u_i \in u,c$), also include (by virtue of gauge invariance) the charged 4-Fermi currents $t d \ell \nu_\ell$ ($d \in d,s$). As such, these operators will also lead to the single-top + single-lepton signals with 0 and 1 accompanying light jet and missing energy, in analogy to the dilepton signals of \eqref{eq:llt}: 
\begin{eqnarray}
&&pp \to \ell + t + \missET~  \nonumber \\
&&pp \to \ell + t  + j + \missET ~,
\label{eq:lt}
\end{eqnarray}
where the underlying production mechanisms for these processes 
are similar to the ones depicted in Fig.~\ref{fig:Feynman}, 
replacing $u \to d$ and one of the charged leptons with a neutrino in these diagrams\footnote{Note that another related operator 
${\cal O}_{ledq}=(\bar\ell e)(\bar d q)$, which is not considered in this work (i.e., since it does not yield the FC $tu_i \ell \ell$ interactions that lead to our single-top + dilepton signals) 
can also contribute to the mono-top + single-lepton signals in \eqref{eq:lt}.}. 

Similarly, the $(\bar LL)(\bar L L)$ operators 
${\cal O}_{l q}^{(1)}, {\cal O}_{l q}^{(3)}$ and 
the $(\bar LL)(\bar R R)$ operator ${\cal O}_{l u}$ 
with flavor indices contributing to the 
$t u_i \ell \ell$ interactions, also generate 
(again, by virtue of gauge invariance) the FCNC 4-Fermi terms involving only neutrinos, i.e., 
$t u_i \nu_\ell \nu_\ell$. They therefore also lead to 
the following signals of a single-top + $\missET$ with and without a light-jet (and without charged leptons):
\begin{eqnarray}
&&pp \to t + \missET~  \nonumber \\
&&pp \to t  + j + \missET ~,
\label{eq:t}
\end{eqnarray}
where, here also, the underlying diagrams for these processes are similar to the ones depicted in Fig.~\ref{fig:Feynman}, replacing the two charged leptons with two neutrinos.

These single-top + single (charged) lepton + $\missET$ and single-top + $\missET$ signals are more challenging, since they are expected to have a significantly larger SM background, e.g., from $t \bar t$ production. Nonetheless, a dedicated study of such beyond the SM signals is called for, where the  correlations between the no-lepton, mono-lepton and di-lepton single-top signals can be exploited to gain a better sensitivity, as we have recently demonstrated in~\cite{our_LFU_paper}.   

Following the above discussion, in Table~\ref{tab:processes} we draw a chart which maps the contributions of the
six types of 4-Fermi $tu_i \ell \ell$ operators studied here to the  different types of single-top production processes 
and to $b/B$-physics.  We see that $V_{RR}$ is the only operator which affects only the single-top + dilepton signals studied in this work, without influencing the other single-top channels and $b/B$-physics. 
\begin{table*}[htb]
\caption{Processes affected by the six 4-Fermi $tu_i \ell \ell$ operators type, due to gauge invariance.
$b/B$-physics stands for scattering processes,
not involving top-quarks, with b-quark either in the initial or the final state, and/or NP in $B$-decays. 
See also text. \label{tab:processes}}
\begin{tabular}{c|c|c|c|c|c|c|}
\multirow{2}{*}{} \
 & \multicolumn{6}{c|}{$tu_i \ell \ell$ 4-Fermi type} 
\tabularnewline
\cline{2-7} 
 signal & $S_{RR}$ & $T_{RR}$ & $V_{RR}$ & $V_{LL}$ &
 $V_{RL}$ & $V_{LR}$
\tabularnewline
\hline 
\
$ pp \to t \ell \ell$/$t \ell \ell +j$
 & $\checkmark$ & $\checkmark$ & $\checkmark$ & $\checkmark$ &$\checkmark$ & $\checkmark$ \tabularnewline
\hline 
\
$ pp \to t \ell + \missET$/$t \ell +j +\missET$
 & $\checkmark$ & $\checkmark$ &  & $\checkmark$ &&  \tabularnewline
\hline 
\
$ pp \to t +\missET$/$t +j + \missET$
 &  & &  & $\checkmark$ & & $\checkmark$ \tabularnewline
\hline 
\
$b/B$-physics 
 & $\checkmark$ & $\checkmark$ &  & $\checkmark$ &$\checkmark$ &  \tabularnewline
\hline 
\end{tabular}
\end{table*}

\section{Signal and background analysis}

In this section we will describe the essential ingredients for the signal over background analysis of the single top + di-lepton signal. Specifically, we will provide a sensitivity study to the NP signals, based on simplified criteria. A more realistic analysis will be presented in the next section.      

We will use an $m_{\ell \ell}$-dependent integrated cross-section, selecting events above a minimum value of $m_{\ell\ell}$:
\begin{eqnarray}
\sigma(m_{\ell \ell}^{\tt min}) \equiv 
\sigma( m_{\ell \ell} \geq m_{\ell \ell}^{\tt min}) = 
\int_{m_{\ell \ell} \geq m_{\ell \ell}^{\tt min}} d m_{\ell \ell} \frac{d\sigma}{dm_{\ell \ell}} ~, \label{CCSX}
\end{eqnarray}
where $m_{\ell \ell}^{\tt min}$ will be chosen to optimize the analysis  sensitivity. In the next section we will also impose an upper cut, $m_{\ell \ell} \leq m_{\ell \ell}^{\tt max}$, that will be used to ensure the applicability of the EFT approach we adopt.

The cross-section for the single-top + di-lepton production channels in \eqref{eq:llt} can then be written in the general form
\begin{eqnarray}
\sigma_{t \ell \ell_j}(m_{\ell \ell}^{\tt min})=\sigma_{t \ell \ell_j}^{{\tt SM}}(m_{\ell \ell}^{\tt min}) + \frac{f^2}{\left(\Lambda/\left[{\tt TeV} \right]\right)^4} \cdot \sigma_{t \ell \ell_j}^{{\tt NP}}(m_{\ell \ell}^{\tt min}) ~, \label{EFT_exp}
\end{eqnarray}
where $\sigma^{\tt SM}_{t \ell \ell_j}(m_{\ell \ell}^{\tt min})$ and $\sigma^{\tt NP}_{t \ell \ell_j}(m_{\ell \ell}^{\tt min})$  are the $m_{\ell \ell}$-dependent SM and NP$^2$ integrated cross-sections, respectively. 
We recall that 
$\sigma_{t \ell \ell_0}^{{\tt SM}} =0$ at tree-level and that the 1-loop contribution is vanishingly small (see the discussion in section~\ref{sec:intro}). Furthermore, 
the NP diagrams are QCD-generated (via gluon-quark and gluon-gluon fusion, see Fig.~\ref{fig:Feynman}) and, in the $t \ell \ell_1$ case, they do not interfere with the SM, which is electroweak-generated and involves a different final state 
(see Fig.~\ref{fig:SM}).
Therefore, there is no term $ \propto 1/\Lambda^2 $ for both the $t \ell \ell_0$  and $t \ell \ell_1$ channels, so that the leading NP terms are scaled by the NP couplings $f^2/\Lambda^{4}$, where $f$ is the dimensionless coefficient of the 4-Fermi $tu_i \ell \ell$ interactions in \eqref{4fermimatrix} and \eqref{vst}, i.e., $f= V_{ij},S_{ij}$ or $T_{ij}$ for the vector, scalar or tensor-like terms and we will take $f=1$ henceforward for simplicity.

All cross-sections reported in this section were calculated using {\sc MadGraph5\_aMC@NLO} \cite{madgraph5} at LO parton-level and a dedicated universal FeynRules output (UFO) model for the EFT framework was produced using {\sc FeynRules}~\cite{FRpaper}. We have used the LO MSTW 2008 parton distribution functions (PDF) set (MSTW2008lo68cl~\cite{MSTW2008}) in the 5-flavor scheme  with a dynamical scale choice for the central value of the factorization ($\mu_F$) and renormalization ($\mu_R$) scales, i.e., corresponding to the sum of the transverse mass in the hard-processes.  As a baseline selection, we used: $p_T(j) >  35$~GeV, $|\eta(j)| < 4.5$ for jets and $p_T(\ell) > 25$~GeV,$|\eta(\ell)| < 2.5$ for leptons. Also, the minimum angular distance in the $\eta-\phi$ plane between all objects (leptons and jets) is $> 0.4$ and kinematic selections cuts (i.e., on the di-lepton invariant mass) were imposed using {\sc MadAnalysis5}~\cite{madanalysis5}.

\begin{figure}[htb]
\includegraphics[width=0.45\textwidth]{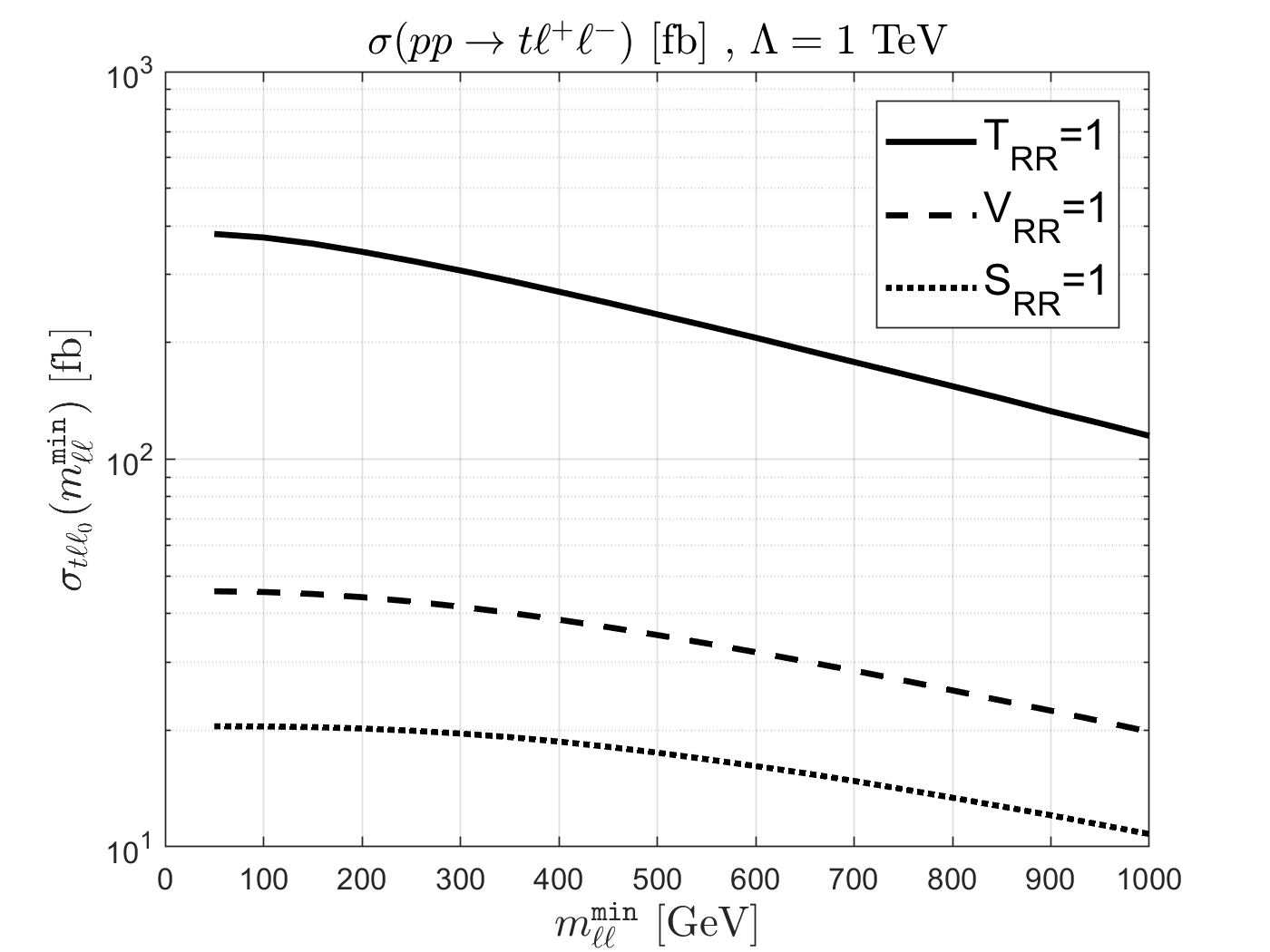}
\includegraphics[width=0.45\textwidth]{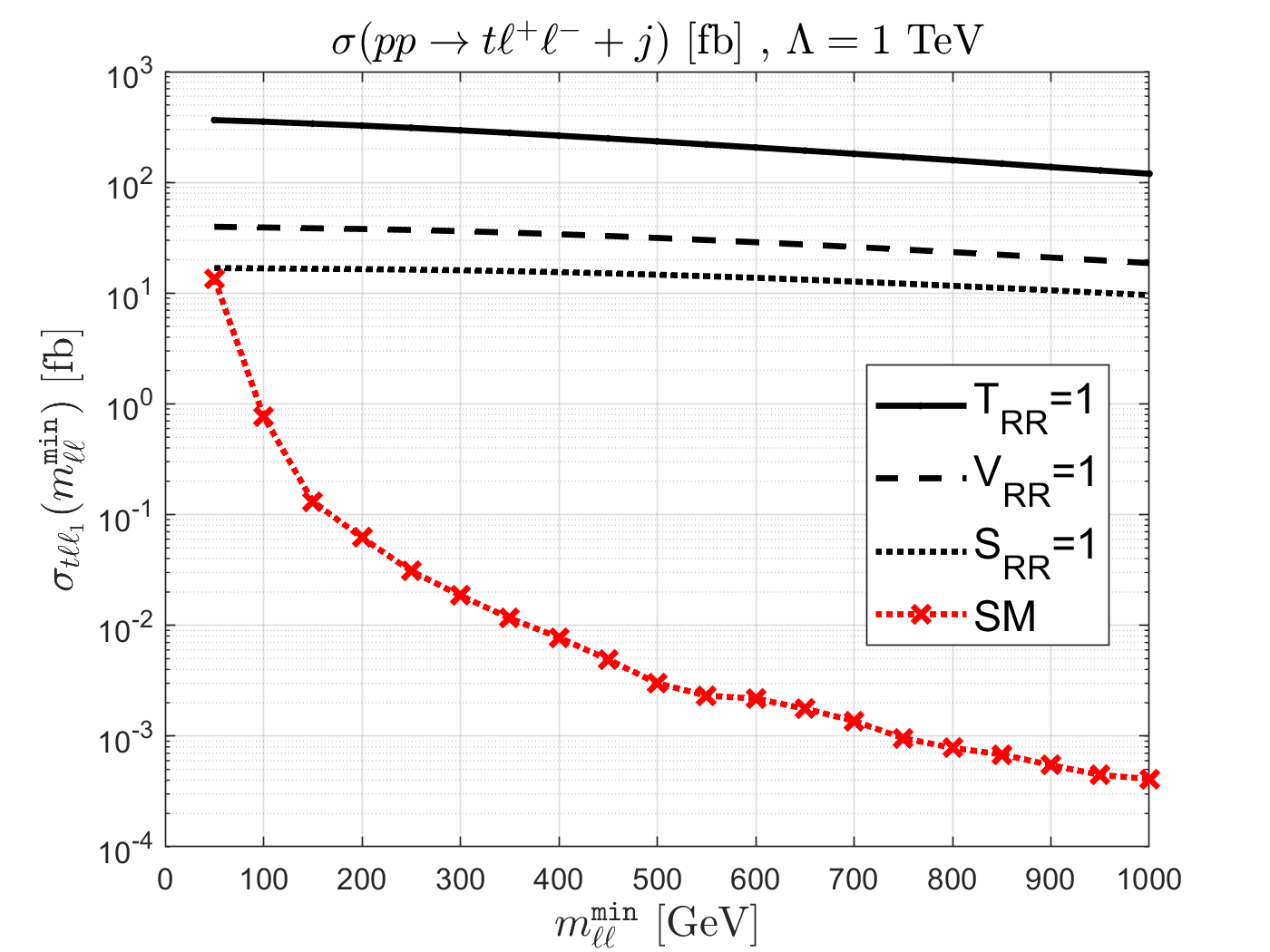}
\caption{Integrated cross-sections for the $(t \ell \ell)_0$ and 
$(t \ell \ell)_1$ single-top + di-lepton channels, as a function of the lower di-lepton invariant mass cut, see \eqref{EFT_exp}. The NP effects are calculated with $\Lambda=1$~TeV and $f=1$, 
where $f= V_{RR},f=S_{RR}$ or $f=T_{RR}$.}
\label{fig:tllCSX}
\end{figure}

To get an estimate of the sensitivity of the results to the lower cut selection of the di-lepton invariant mass $m_{\ell \ell}^{\tt min}$, we plot in Fig.~\ref{fig:tllCSX} and list in Table~\ref{tab:tllCSX} the NP and SM   integrated $t \ell \ell_j$ cross-sections, as a function of $m_{\ell \ell}^{\tt min}$ (at this point without an upper cut selection $m_{\ell \ell}^{\tt max}$), where the NP terms were calculated for the scalar, vector and tensor 4-Fermi operators with the benchmark values of $\Lambda=1$~TeV and $f=1$ ($f= S_{RR},~V_{RR},~T_{RR}$). Results for different choice of $ \Lambda$ and/or $f$ are obtained by scaling the cross-section by $ f^2/\Lambda^4$. 
We see that selecting high $m_{\ell \ell}$ di-lepton events, $m_{\ell \ell}^{\tt min} > 100$~GeV,  
the SM contribution for the $(t \ell \ell)_1$ is dramatically suppressed (i.e., by about five orders of magnitude in going from $m_{\ell \ell}^{\tt min}=50$~GeV to $m_{\ell \ell}^{\tt min}=1$~TeV). With this choice we have $\sigma_{t \ell \ell_1}^{{\tt SM}} \ll \sigma_{t \ell \ell_1}^{{\tt NP}}$, so that the SM contribution can also be ignored in \eqref{EFT_exp}. Indeed,  as shown below, we obtain a better sensitivity to the NP with a higher $m_{\ell \ell}^{\tt min}$ selection, for which, not only the SM irreducible background effectively vanishes, but also the potential reducible SM background is essentially eliminated.

In the following we will study the sensitivity only to the $S_{RR},V_{RR}$ and $T_{RR}$ 4-Fermi interactions, noting that, since we are mainly analysing total cross-sections and since there are no SM$\times$NP interference effects (see discussion above), the sensitivity and reach for the other 4-Fermi vector currents, $V_{LL,},V_{RL}$ and $V_{LR}$ is identical to that of $V_{RR}$.  

\begin{table*}[htb]
\caption{The integrated cross-sections of 
the pure NP contributions to the processes $(t \ell \ell)_0$ and $(t \ell \ell)_1$ of \eqref{eq:llt}, i.e., $\sigma_{t \ell \ell_0}^{{\tt NP}}$ and $\sigma_{t \ell \ell_1}^{{\tt NP}}$ in \eqref{EFT_exp}, and of the SM part in the $(t \ell \ell)_1$ channel, $\sigma_{t \ell \ell_1}^{{\tt SM}}$ in \eqref{EFT_exp}, for the di-lepton invariant mass lower cut selections $m_{\ell \ell}^{\tt min}=50,\,100,\,300,\,1000$~GeV.
The NP contributions are calculated with 
$\Lambda=1$~TeV and $f=1$, where $f=S_{RR},T_{RR}$ or
$f=V_{RR}$. See also text. \label{tab:tllCSX}}
\begin{tabular}{c|c|c|c|c|}
\multirow{2}{*}{} \
 & \multicolumn{4}{c|}{Integrated cross-section [fb], $\Lambda =1$~TeV} 
\tabularnewline
\cline{2-5} 
 Source & $m_{\mu^+ \mu^-}^{\tt min}=50$~GeV & $m_{\mu^+ \mu^-}^{\tt min}=100$~GeV & $m_{\mu^+ \mu^-}^{\tt min}=300$~GeV & $m_{\mu^+ \mu^-}^{\tt min}=1000$~GeV 
\tabularnewline
\hline 
\
NP: $pp \to t \ell^+ \ell^-$ ($S_{RR}=1$)
 & 20.5 & 20.4 & 19.6 & 10.8  \tabularnewline
\hline 
\
NP: $pp \to t \ell^+ \ell^-$ ($T_{RR}=1$)
 & 381.2 & 373.3 & 306.9 & 114.9  \tabularnewline
\hline 
\
NP: $pp \to t \ell^+ \ell^-$ ($V_{RR}=1$)
 & 45.6 & 45.5 & 41.6 & 19.8  \tabularnewline
\hline 
\
NP: $pp \to t \ell^+ \ell^- j$ ($S_{RR}=1$)
& 16.8 & 16.7 & 16.1 & 9.6
\tabularnewline
\hline 
\
NP: $pp \to t \ell^+ \ell^- j$ ($T_{RR}=1$)
& 365.0 & 353.2 & 295.4 & 119.6
\tabularnewline
\hline 
\
NP: $pp \to t \ell^+ \ell^- j$ ($V_{RR}=1$)
& 39.7 & 39.2 & 36.3 & 18.7
\tabularnewline
\hline 
\
SM: $pp \to t \ell^+ \ell^- j$ & 13.6 & 0.77 & 0.019 & 0.00041  \tabularnewline
\hline 
\end{tabular}
\end{table*}

\subsection{Event selection: signal vs. background}

To study the sensitivity to the NP, we will isolate the signal using either an inclusive tri-lepton selection criteria or a di-lepton signature with an additional selection of a single $b$-tagged jet:\footnote{Although top reconstruction will not be considered here, it may be useful for further reducing the background, e.g., the  $VV$ and $Z$+jets backgrounds considered below.}
\begin{eqnarray}
(\ell^\prime \ell \ell) &:& 
pp \to \ell^\prime \ell^+ \ell^- + X ~, 
\label{eq:lll_inclusive} \\
(\ell \ell 1b) &:& 
pp \to \ell^+ \ell^- + j_b + X ~ ,
\label{eq:lljb}
\end{eqnarray}
so that in the tri-lepton case we select events where the top decays via $t \to b W \to b \ell^\prime \nu_{\ell^\prime}$ and demand exactly 3 (isolated) charged leptons in the final state, where, in general, $\ell,\ell^\prime = e,\mu$ or $\tau$ and $\ell^\prime = \ell$ and/or $\ell^\prime \ne \ell$ can be considered.
An additional selection of a single $b$-tagged jet with the tri-lepton signal, i.e., $(\ell^\prime \ell \ell 1b): pp \to \ell^\prime \ell^+ \ell^- + j_b + X$ may in some cases also improve the sensitivity to the scale of the $t u_i \ell \ell$ 4-Fermi operators; we will briefly comment on that in the next section. 
We note that the $(\ell^\prime \ell \ell)$ tri-lepton selection was recently used by both ATLAS~\cite{Aaboud:2017ylb,Aad:2020wog} and CMS~\cite{Sirunyan:2017nbr,Sirunyan:2018zgs} in the measurement of the SM $p p \to t \ell^+ \ell^- j$ cross-section (see also~\cite{Campbell:2013yla,tZ1,tZ8-trilepton,Sirunyan:2020tqm}). 
In fact, these tri-lepton signatures (with or without a high-$p_T$ jet; either light-jet or a $b$-jet) are rich in phenomenology, as they can probe several types of other well motivated~TeV-scale NP scenarios, e.g., electroweak pair production of charginos and neutralinos in supersymmetry~\cite{ATLAS_3l} and the production of a heavy neutral Majorana-type lepton
\cite{CMS_heavy_lepton_3l}.

We find that the tri-lepton $(\ell^\prime \ell \ell)$ or di-lepton $(\ell \ell 1b)$ signal selections of \eqref{eq:lll_inclusive} and \eqref{eq:lljb} 
along with the selection of events with high $\ell^+ \ell^-$ invariant mass are sufficient to reduce the potential SM background to the level that it can be neglected. Furthermore, selecting a single $b$-tagged jet is found to be crucial in the case of the di-lepton signal for efficiently tagging the top-quark decay in the final state and isolating the signal from the background 
(see also~\cite{bsll,bbll,Marzocca:2020ueu}, where a better sensitivity for NP effects in di-lepton events was obtained with a single $b$-tagged jet selection in $pp \to \mu^+ \mu^- + j_b$ and $pp \to \tau \nu_\tau + j_b$).

As a case study, for the rest of this section we will assume that the NP  generates only the di-muon 4-Fermi interactions and focus below either on the $(e \mu \mu)$ tri-lepton channel $pp \to e^\pm \mu^+ \mu^- +X$ or the $(\mu \mu 1b)$ signal 
$pp \to \mu^+ \mu^- +j_b +X$. We note, though, that similar analyses can be performed for the  tri-lepton case in the channels $e \mu \mu, \mu \mu \mu, \tau \mu \mu$ or, more generally, for the channels $e \ell \ell, \mu \ell \ell, \tau \ell \ell$ when the NP generates the $t u_i \ell \ell$ operators for any given lepton flavor. Tri-lepton final states with 3 identical leptons can be similarly analysed with appropriate selections on any pair of OSSF leptons. 
However, both the di-lepton and tri-lepton final states involving the $\tau$ are more challenging and are expected to have a decreased sensitivity due to the lower experimental detection efficiency for the $\tau$. 

The leading  
potential background for the $t \mu \mu$ and $t \mu \mu j$ signals arise 
from the SM $t \bar t$ and $\mu^+\mu^-$+ jets (dubbed hereafter as $Z$+jets) production channels:
\begin{itemize}
\item $t \bar t$: $pp \to t \bar t$, followed by leptonic top-quark decays to muons $t \bar t \to \mu^+ \mu^- + 2 j_b + \missET$
\item $Z$+jets: $pp \to \mu^+ \mu^- + {\tt jets}$
\end{itemize}
which pass the tri-lepton selection when a non-prompt or fake lepton originate from hadronic decays or from mis-identified jets. 
Additional sources of background, which we find to be sub-leading (see Table \ref{tab:emumu_Yields} in the next section), include the
$VV$, $tW$, $t \bar t V$ and $t \bar VV$ production channels, 
where $V=W,Z,\gamma$. For example, for the $(e \mu \mu)$ tri-lepton signal these are:
\begin{itemize}
\item $WZ$: $pp \to W \mu^+ \mu^-$, followed by $W \to e \nu_e$
\item $ZZ$: $pp \to Z \mu^+ \mu^-$, followed by $Z \to e^+ e^-$ (contributes in case one electron is not tracked)
\item $tW$: $pp \to t W$, followed e.g., by $t W \to \mu^\pm \mu^\mp + j_b  + 2j + \missET$ (+ a non-prompt electron, see text)
\item $t \bar t W$: $pp \to t \bar t \mu^\pm \nu_\mu$, followed 
by the leptonic top decays 
$t \bar t \to e^\pm \mu^\mp + 2 j_b  + \missET$ \\
$~~~~~~~~pp \to t \bar t e^\pm \nu_e$, followed by
$t \bar t \to \mu^+ \mu^- + 2 j_b  + \missET$
\item $t \bar t Z$: $pp \to t \bar t \mu^+ \mu^-$,  followed by the top decays $t \bar t \to e^\pm + 2 j_b  + 2j + \missET$ 
\item $t \bar WZ$: $pp \to t W Z$, followed by $Z \to \mu^+ \mu^-$ and 
$t W \to e^\pm + j_b  + 2j + \missET$
\end{itemize}

In Tables~\ref{tab:emumu_signal_tull} we list the number of inclusive 
tri-lepton $pp \to e \mu^+ \mu^-$ events 
per 100 fb$^{-1}$ of integrated luminosity, with the selections $m_{\mu^+ \mu^-}^{\tt min}=100,\,300,\,500,\,1000$~GeV,   
generated by the $tu \mu \mu$ and $t c \mu \mu$ 4-Fermi operators and from the irreducible SM process $pp \to t \mu^+ \mu^- j$. Note that the corresponding number of 
$pp \to \mu^+ \mu^- + j_b$ events are 9 times larger than the number of inclusive 
tri-lepton $e \mu^+ \mu^-$ events listed in Tables~\ref{tab:emumu_signal_tull}, since all top decay channels $t \to bW$ followed by both the leptonic and the hadronic $W$-decays
are included in this case.
As discussed above, the $m_{\mu^+ \mu^-}^{\tt min}$ selection is very effective for reducing the background to both the $(\mu \mu 1b)$  and $(e \mu\mu)$ signals; see Tables \ref{tab:mumu_Yields} and \ref{tab:emumu_Yields} in the next section, where we list the yields from various sources of backgrounds to the $(\mu \mu 1b)$ and $(e \mu\mu)$ signals for the selections $m_{\mu^+ \mu^-}^{\tt min}=500,\,1000,\,1500,\,2000$~GeV. For example, the background to the 
inclusive $(e \mu\mu)$ tri-lepton signal becomes negligible with the selection of 
$m_{\mu^+ \mu^-}^{\tt min} \sim 1000$~GeV. Thus, to get an estimate of the sensitivity to the new $t u_i \mu \mu$ 4-Fermi operators, we will consider in the rest of this section only the  inclusive $e\mu\mu$ tri-lepton signal case with $m_{\mu^+ \mu^-}^{\tt min} = 1000$~GeV and assume 
that it is background free in this regime. 
A more realistic analysis including both the tri-lepton and di-lepton + $b$-jet selections will be presented in the next section.
\begin{table*}[htb]
\caption{Number of $(e \mu \mu)$ signal events per 100 fb$^{-1}$ of integrated luminosity, expected from the irreducible SM process $pp \to t \mu^+ \mu^- j$ and from the 
pure NP $tu \mu \mu$ and $tc \mu \mu$ contributions to the fully inclusive 
$pp \to e \mu \mu +X$ signal as defined in \eqref{eq:lll_inclusive}, 
with dimuon invariant mass lower cut selections of $m_{\mu^+ \mu^-}^{\tt min}=100,300,500,1000$~GeV.  
The NP contributions are calculated with 
$\Lambda=1$~TeV and $f=1$, where $f=S_{RR},T_{RR}$ or
$f=V_{RR}$. 
Note that the number of inclusive NP events includes the contributions from both $pp \to t \mu^+ \mu^-$ and $pp \to t \mu^+ \mu^- j$, followed by $t \to b e \nu_e$. 
See also text. \label{tab:emumu_signal_tull}}
\begin{tabular}{c|c|c|c|c|c|}
\multirow{1}{*}{} \
  & \multicolumn{5}{c|}{Number of inclusive $pp \to e \mu^+ \mu^- + X$ signal events/100 fb$^{-1}$, $\Lambda=1$~TeV} 
\tabularnewline
\cline{2-6} 
 Source & Coupling & $m_{\mu^+ \mu^-}^{\tt min}=100$~GeV & $m_{\mu^+ \mu^-}^{\tt min}=300$~GeV & $m_{\mu^+ \mu^-}^{\tt min}=500$~GeV & $m_{\mu^+ \mu^-}^{\tt min}=1000$~GeV 
\tabularnewline
\hline \hline
\
  & $S_{RR}=1$ & 399 & 382 & 342 & 215 \tabularnewline
\cline{2-6} 
\
\boldmath{$tu \mu\mu$} {\bf 4-Fermi} & $T_{RR}=1$ & 7937 & 6568 & 5117 & 2539  \tabularnewline
\cline{2-6} 
\
& $V_{RR}=1$
 & 916 & 841 & 716 & 409   \tabularnewline
\hline 
\hline 
  & $S_{RR}=1$ & 29 & 25 & 20 & 9 \tabularnewline
\cline{2-6} 
\
\boldmath{$tc \mu\mu$} {\bf 4-Fermi} & $T_{RR}=1$ & 711 & 481 & 318 & 108  \tabularnewline
\cline{2-6} 
\
& $V_{RR}=1$
 & 75 & 60 & 44 & 18   \tabularnewline
\hline 
\hline 
\multirow{1}{*}{$pp \to t \mu^+ \mu^- j$} & SM irreducible & 8 & 0 & 0 & 0 \tabularnewline
\hline 
\hline 
\end{tabular}
\end{table*}

\subsection{Domain of validity of the EFT setup \label{EFT_validity}}
The basic assumption underlying the EFT approach is that none of the heavy particles can be directly produced in the processes being investigated. Assuming that $ \Lambda $ represents the masses of these particles, this leads to the requirement $ \Lambda^2 \gtrsim \hat s$, where $\sqrt{\hat s}$ is the center-of-mass energy of the hard process. Alternatively, it is required that the NP cross-sections do not violate tree-level unitarity bounds, which  leads to similar constraints (for the case at hand the FC $tu_i \ell \ell$ 4-Fermi operators generate a cross-section that grows with energy $\sigma^{\tt NP}_{t \ell \ell_j} \propto \hat s$). These criteria, however, are not precise enough for our purposes for the following reasons:
\begin{itemize}
    \item The 4-fermion operators we consider can be generated either by a $Z$-like heavy particle coupling to lepton and quark pairs ({\it eg.} $t u_i \to X \to \ell \ell$), or by a leptoquark coupling to quark-lepton pairs ({\it eg.} $t \ell \to {\tt LQ} \to u_i \ell$). In the first case the EFT is applicable when $\Lambda > m_{\ell\ell}^{\tt max}$, and in the second case when $\Lambda > m_{q\ell}^{\tt max}$. If only one of these conditions is obeyed the EFT approach remains applicable, but only for the corresponding type of new physics; only when {\em both} are violated is the EFT approach unreliable. It is worth noting that $\Lambda > m_{q\ell}^{\tt max}$ is often much less restrictive.
    
    \item The constraints we derive will be on the effective scale $\Lambda_{\tt eff} = \Lambda/\sqrt{f}$, whence the EFT applicability conditions become $\Lambda_{\tt eff} >  m_{q\ell,\, \ell\ell}^{\tt max}/\sqrt{f}$. Thus, the EFT approach remains applicable even for situations where $\Lambda_{\tt eff}$ is of the same order, or even somewhat smaller\footnote{Note for example that applying the naive EFT validity criteria, $s < \Lambda_{\tt eff}^2$, to the Fermi theory of weak interactions would give $s < (246~{\rm GeV})^2$ if $f=1$, but in reality $f \sim 0.3 $ and therefore $s \lsim (100~{\rm GeV})^2$.} than $m_{q\ell,\,\ell\ell}^{\tt max}$. This corresponds to NP scenarios with $ f>1$ (while still remaining perturbative).
\end{itemize}

Based on this we will define the region of applicability by demanding $ \Lambda > m_{\ell\ell}^{\tt max}$ {\em or} $ \Lambda > m_{q\ell}^{\tt max}$, and allow $ \Lambda_{\tt eff} $ to be smaller than $m_{q\ell,\, \ell\ell}^{\tt max}$ by an $O(1)$ factor.

To close this section we note that dimension 8 operators that interfere with the SM also generate ${\cal O}(\Lambda^{-4})$ contributions to the $pp \to t \mu^+ \mu^- j$ cross section. These, however, can be ignored compared to the ${\cal O}(\Lambda^{-4})$ NP(dim.6)$\times$NP(dim.6) terms that we keep, because the SM amplitude is much suppressed for the high $m_{\ell \ell}^{\tt min}$ selection that we use, as noted above.

\subsection{Sensitivity to the NP}

We have not considered up to this point the theoretical and experimental uncertainties involved with the calculation and measurement of our $pp \to t \ell^+ \ell^- + X \to \ell^\prime \ell^+ \ell^- +X$  signals. The theoretical uncertainties are due to the flavor scheme (i.e., 4-flavor vs. 5-flavor), the NLO QCD (K-factor) and EW corrections, the dependence on the renormalization ($\mu_R$) and factorization ($\mu_F$) scales and the uncertainty due to the PDF choice. A detailed study of the SM dilepton + single-top associated production 
$pp \to t \ell^+ \ell^- j$ was recently performed in~\cite{Pagani:2020mov}, 
where it was found that these theory uncertainties amount to an ${\cal O}(10\%)$ uncertainty
in the estimate of $\sigma(pp \to t \ell^+ \ell^- j)$. 
It should be noted, though, that the uncertainties reported in \cite{Pagani:2020mov} may not necessarily apply to out study, since our dominant signal processes are different (different initial and final states) and, also, we focus on a different kinematical region of the di-lepton signals: the high di-lepton invariant mass part of the phase-space, $m_{\ell^+ \ell^-} > 1$~TeV
(see below and in the next section).

\begin{figure*}[htb]
\centering
\includegraphics[width=0.45\textwidth]{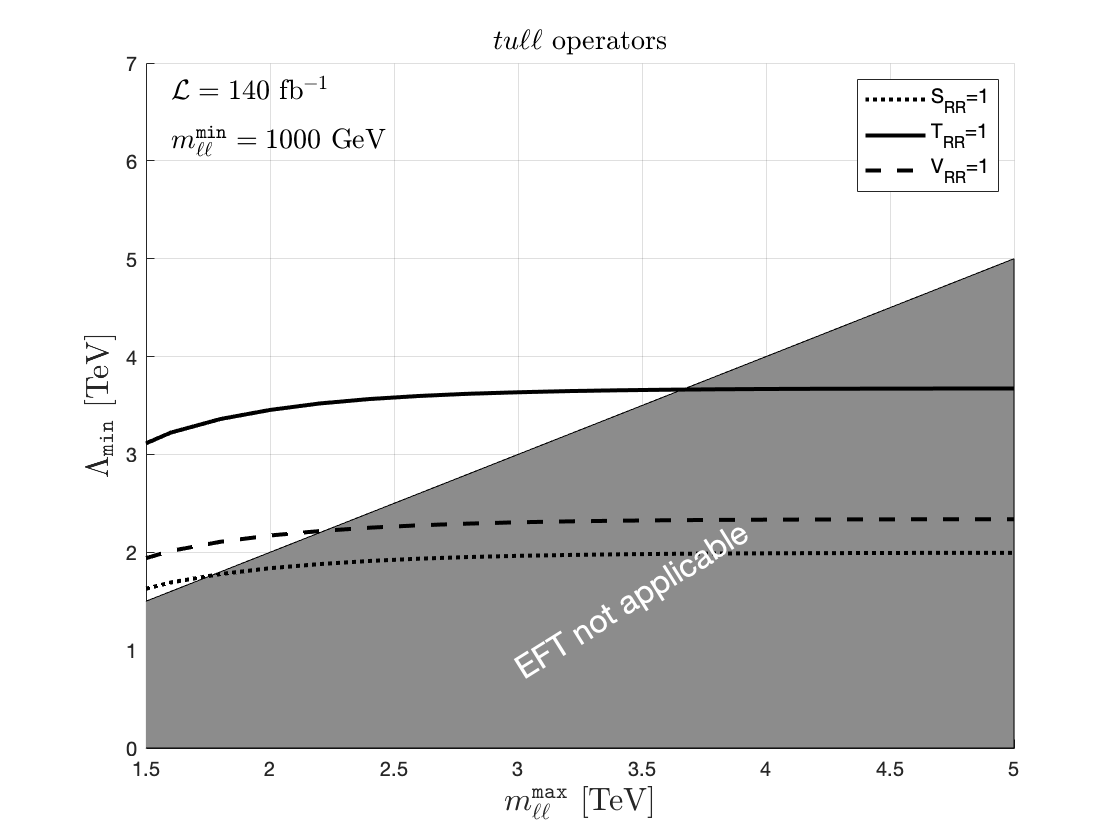}
\includegraphics[width=0.45\textwidth]{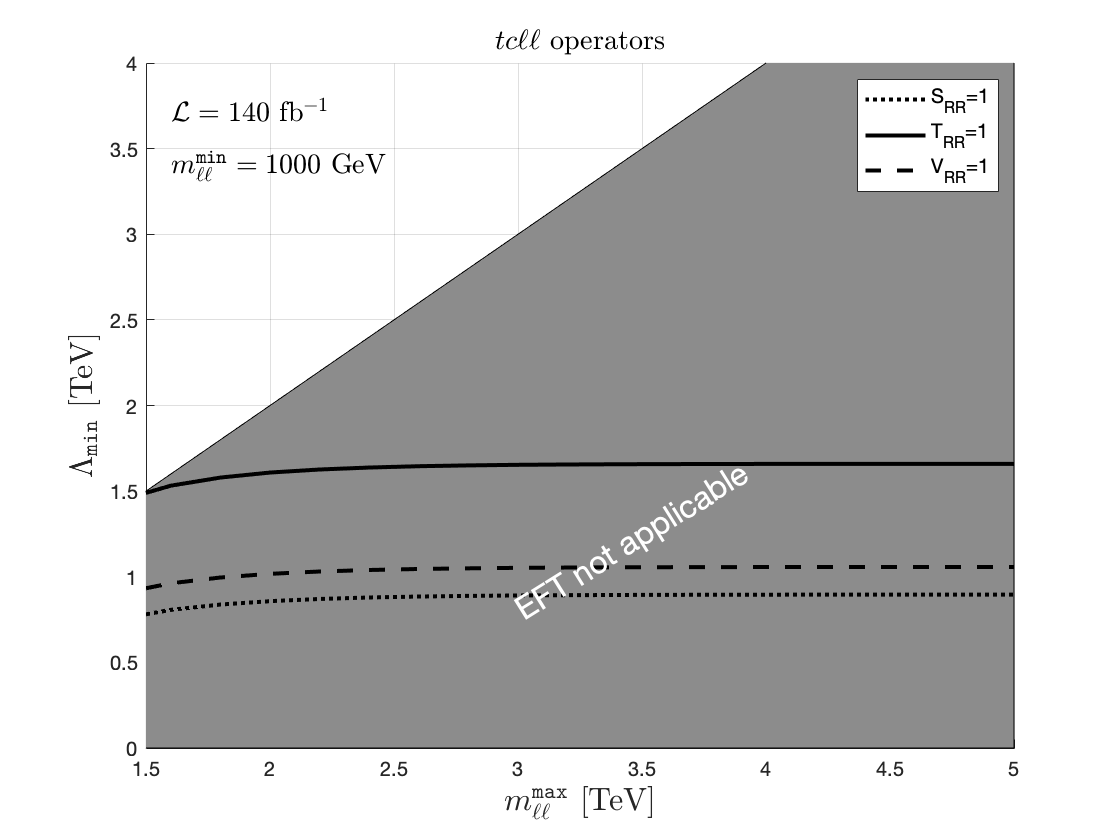}
\caption{Expected bounds ($\Lambda_{\tt min}$) on the scale $\Lambda$ (i.e., $\Lambda > \Lambda_{\tt min}$) from the tri-lepton signal $pp \to e \mu^+ \mu^- + X$, for the scalar (dotted-line), vector (dashed-line) and tensor (solid-line) 4-Fermi $t u \mu \mu$ (left) and $tc \mu \mu$ (right) operators with $S_{RR}=1$, $T_{RR}=1$ and $V_{RR}=1$, as a function of an upper cut on the di-muon invariant mass $m_{\mu^+ \mu^-}^{\tt max}$.  
The bounds are calculated with $m_{\mu^+ \mu^-}^{\tt min}=1000$~GeV and for an integrated luminosity of ${\cal L}=140$ fb$^{-1}$. 
The shaded areas correspond to the region where
$m_{\mu^+ \mu^-}^{\tt max} > \Lambda_{\tt min}$, which naively represents the domain outside the validity of the EFT prescription. See also text.
}
\centering
\label{fig:EFT_V1}
\end{figure*}

Thus, the overall experimental uncertainty (i.e., statistical and systematic) for our inclusive tri-lepton signals is not known, in particular, in the high di-lepton invariant mass regime (the sensitivity of our results to the overall uncertainty will be examine in the next section). Thus, to be on the conservative side, for an estimate of the sensitivity to the scale of the $tu_i \mu \mu$ 4-Fermi operators, we demand at least 20 inclusive $(e \mu \mu)$ tri-lepton  events with the high $m_{\mu \mu}^{\tt min}=1000$~GeV selection, which ensures at least 10 background-free NP 
events (see discussion above) even with an overall theoretical + experimental uncertainty of ${\cal O}(50\%)$.
This condition corresponds to:
\begin{eqnarray}
\frac{f^2}{\left(\Lambda/\left[{\tt~TeV}\right]\right)^4} \cdot \sigma_{e \mu \mu}^{{\tt NP}}(m_{\mu \mu}^{\tt min}) \cdot {\cal L} \ge 20 \label{bounds}~,
\end{eqnarray}
where 

\begin{eqnarray}
\sigma_{e \mu \mu}^{{\tt NP}}(m_{\mu \mu}^{\tt min}) = \left[ \sigma_{t \mu \mu_0}^{{\tt NP}}(m_{\mu \mu}^{\tt min}) + \sigma_{t \mu \mu_1}^{{\tt NP}}(m_{\mu \mu}^{\tt min}) \right] \cdot {\tt BR}(t \to b e \nu_e) \label{bounds_2}~,
\end{eqnarray}
and $\sigma_{t \mu \mu_j}^{{\tt NP}}(m_{\mu \mu}^{\tt min})$ with $j=0,1$ are the NP parts of the integrated cross-sections in \eqref{EFT_exp} 
for the $(t\mu \mu)_0$ and $(t\mu \mu)_1$ single-top signals of \eqref{eq:llt}.

The requirement \eqref{bounds} must be complemented with the constraints imposed by the validity of the EFT approach, $ \Lambda > m^{\tt max}_{\ell\ell}$, as discussed in Sect. \ref{EFT_validity} above. As an example, we take $ S_{RR} = V_{RR} = T_{RR}=1$ (corresponding to $ f =1 $ and $ \Lambda = \Lambda_{\tt eff}$) and 
plot in Fig.~\ref{fig:EFT_V1} the expected bounds on $\Lambda$ ($\Lambda_{\tt min}$) of the scalar, vector and tensor $tu \mu \mu$ and $tc \mu \mu$ operators, as a function of the upper cut $m^{\tt max}_{\ell\ell}$, for an integrated luminosity of ${\cal L}=140$ fb$^{-1}$ and a lower cut selection 
of $m_{\mu^+ \mu^-}^{\tt min}=1000$~GeV on the di-muon invariant mass. 
The shaded regions in Fig.~\ref{fig:EFT_V1} correspond 
to $m^{\tt max}_{\ell\ell} > \Lambda$, which as discussed above, is the domain where the validity of the EFT might be questionable.
The corresponding bounds for integrated luminosities of ${\cal L}=140$ and $3000$ fb$^{-1}$ are summarized in Table~\ref{tab:bounds_tull}, where we note by \noeft~ the cases where the bound is not consistent
with the (conservative) condition of $ \Lambda > m^{\tt max}_{\ell\ell}$ on the applicability of the EFT approach, i.e.,   
the cases where there is no crossing between the curves representing the bounds and the shaded area in Fig.~\ref{fig:EFT_V1}.

We see, for example, that, with the current LHC data of about $140$ fb$^{-1}$, no consistent bound
can be derived on the scale of the $tc\ell \ell$ 4-Fermi operators using the criterion $ \Lambda > m_{\mu^+ \mu^-}^{\tt max} $ for $f=1$, which restricts, but does not necessarily excludes, the EFT approach we adopted as discussed in section \ref{EFT_validity}.

\begin{table*}[htb]
\caption{Expected bounds on the scale of the 4-Fermi
$t u \mu \mu$ and $t c \mu \mu$ operators with $f=1$ ($f=S_{RR},T_{RR},V_{RR}$) from the tri-lepton signal $pp \to e \mu^+ \mu^- + X$ with a di-muon invariant mass lower cut selection of $m_{\mu^+ \mu^-}^{\tt min}=1000$~GeV and for integrated luminosities of ${\cal L}=140$ and $3000$ fb$^{-1}$. Entries marked with \noeft\ refer to cases where $\Lambda < m_{\mu^+ \mu^-}^{\tt max}$, for which the applicability of the EFT approach is questionable. See also text. 
\label{tab:bounds_tull}}
\begin{tabular}{c|c|c|c|}
\multirow{1}{*}{} \
  & \multicolumn{3}{c|}{Expected bounds on $\Lambda$ [TeV], $m_{\mu^+ \mu^-}^{\tt min}=1000$~GeV} 
\tabularnewline
\cline{2-4} 
  & Coupling & \boldmath{$tu \mu\mu$ {\bf 4-Fermi case}} & \boldmath{$tc \mu\mu$ {\bf 4-Fermi case}} 
\tabularnewline
\hline \hline
\
 & $S_{RR}=1$   & 1.8 & 0.9 \noeft \tabularnewline
\cline{2-4} 
\
${\cal L}=140$ fb$^{-1}$ & $T_{RR}=1$  & 3.7 & 1.6 \noeft  \tabularnewline
\cline{2-4} 
\
& $V_{RR}=1$ & 2.2 & 1.1 \noeft \tabularnewline
\hline 
\hline 
\
 & $S_{RR}=1$ & 4.3 & 1.8  \tabularnewline
\cline{2-4}
\
${\cal L}=3000$ fb$^{-1}$ & $T_{RR}=1$ & 7.9 & 3.6  \tabularnewline
\cline{2-4}
\
& $V_{RR}=1$ & 5.0 & 2.2  \tabularnewline
\hline
\hline
\end{tabular}
\end{table*}
\section{A More realistic study}

In order to have a more realistic prediction for the sensitivity to the $tu_i \mu \mu$ 4-Fermi NP terms, we have performed a more detailed analysis for the tri-lepton $(e\mu\mu)$ and di-lepton $(\mu \mu 1b$) signal selections of \eqref{eq:lll_inclusive} and \eqref{eq:lljb} with a pair of OSSF muons: 
$pp \to t \mu^+ \mu^- +X \to e \mu^+ \mu^- + X$ and
$pp \to t \mu^+ \mu^- +X \to \mu^+ \mu^- + j_b + X$, where the electron and $b$-jet originate from the decay of the single-top in the final state.

\subsection{Simulated Event Samples}

All event samples were again generated at LO using {\sc MadGraph5\_aMC@NLO 2.7.3}~\cite{madgraph5} in the 5-flavor scheme, using the dedicated UFO model mentioned earlier that was generated with {\sc FeynRules}~\cite{FRpaper}. Here, the events were then interfaced with the {\sc Pythia 8}~\cite{Mrenna:2016sih} parton shower and
we have used the NNPDF30LO PDF set~\cite{Ball:2014uwa} for samples at $\sqrt{s} = 13$~TeV and the PDF4LHC15 PDF set~\cite{Butterworth:2015oua} for higher centre-of-mass energies (27 and 100~TeV, see below). The 
default {\sc MadGraph5\_aMC@NLO} LO dynamical scale was used, which is the transverse mass calculated by a $k_T$-clustering of the final-state partons.
Events of different jet-multiplicities were matched using the MLM scheme~\cite{MLM} using the default {\sc MadGraph5\_aMC@NLO} parameters and all samples 
were processed through {\sc Delphes 3}~\cite{deFavereau:2013fsa}, which simulates the detector effects, applies simplified reconstruction algorithms and was used for the reconstruction of electrons, muons and hadronic jets. 
For the leptons (electrons and muons) the reconstruction was based on transverse momentum ($p_{\mathrm{T}}$)- and pseudo-rapidity ($\eta$)-dependent artificial efficiency weight and an isolation from other energy-flow objects was applied in a cone of $\Delta R=0.4$ with a minimum $p_{\mathrm{T}}$ requirement of $30$~GeV for each lepton.
Jets were reconstructed using the anti-$k_{t}$~\cite{Cacciari:2008gp} clustering algorithm with radius parameter of $R=0.4$ implemented in FastJet~\cite{Cacciari:2011ma,Cacciari:2005hq}, and were required to have transverse momentum of $p _ {\mathrm{T} } >30$~GeV and pseudo-rapidity $\left|\eta\right|<2.5$.
In cases where a selection of a $b$-jet was used, the identification of $b$-tagged jets was done by applying a $p_ {\mathrm{T}}$-dependent weight based on the jet's associated flavor, 
and the MV2c20 tagging algorithm~\cite{ATL-PHYS-PUB-2015-022} in the 70\% working point.

Several types of background processes were considered (see also discussion in the previous section): (1) the production of 2-3 charged leptons through two gauge bosons (noted as $VV$); (2) the production of 2 muons via a neutral gauge boson (noted as $Z+jets$); (3) the production of two muons from a decay of top-pair (noted as $t \bar t$) and (4) the production of two muons from the decays of top and $W$-boson produced via $pp \to tW$ (noted as $t W$).
For the latter three, an additional non-prompt lepton, which originates e.g., from hadronic decays, can satisfy the tri-lepton selection criterion.
Additional potential background processes (also mentioned earlier) from the SM processes $pp \to t \bar{t}W, t \bar{t}Z, tWZ, t Zq$, were found to be negligible.

\subsection{Event Selection}

As noted above, our base-point selection contains two Opposite-Sign (OS) muons with an additional selection of either one electron in the tri-lepton $pp \to e \mu^+ \mu^- +X$ case, or a single $b$-tagged jet for the di-lepton signal $pp \to \mu^+ \mu^- + j_b + X$.
A requirement of an additional $b$-tagged jet with the tri-lepton signal, 
i.e., the selection $(e \mu \mu 1b)$, will not be considered below, but we note that it can improve the sensitivity obtained with the $(e \mu \mu)$ selection by about 10\% in the high luminosity scenario of the HL-LHC. 

The invariant mass of the OS muons ($m_{\mu^+ \mu^-}$) was used for optimization in both cases, as the discriminating variable between the signal and the background; as shown in the previous section, the NP is expected to dominate at the tail of the $m_{\mu^+ \mu^-}$ distribution whereas a small yield for the SM background is expected in that regime.
We note that a dedicated selection for each signal scenario of the $t u \ell \ell$ or $t c \ell \ell$ operators can improve slightly the sensitivity, but, for simplicity, we keep the selection unified between all three signal scenarios (i.e. $S_{RR},T_{RR},V_{RR}$) of a given operator. 
Furthermore, as mentioned earlier, since the signal contains a single top-quark, a reconstruction of the top quark may also be useful for improving the sensitivity to the NP involved but we will not consider that here. 
Two values of the total integrated luminosity are considered below: 140 and 3000 fb$^{-1}$, which correspond to the currently available and HL-LHC integrated luminosities, respectively. 

\begin{figure}[]
\includegraphics[width=0.49\textwidth]{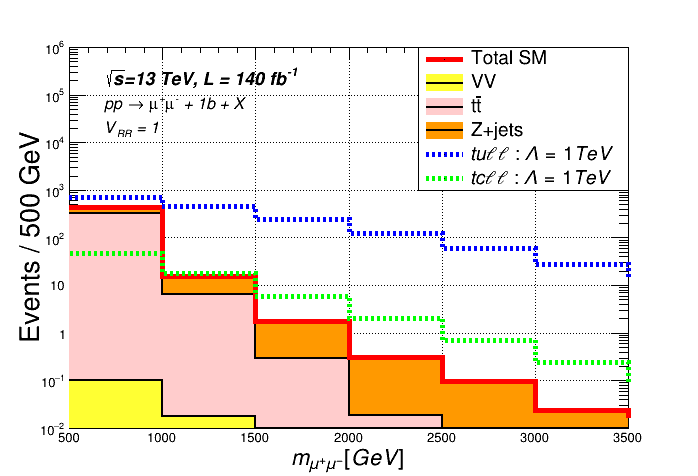}
\includegraphics[width=0.49\textwidth]{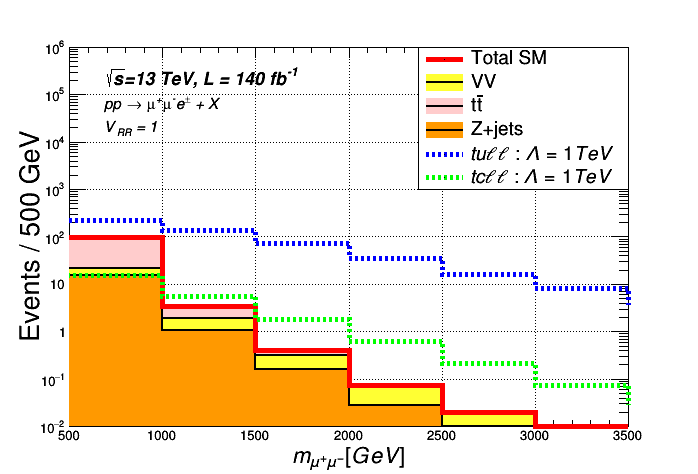}
\caption{
Di-muon invariant mass distribution for the di-lepton $pp \to \mu^+ \mu^- + j_b + X $ (left) and tri-lepton $pp \to e \mu^+ \mu^- +X $ (right) signal scenarios generated by the $t u \ell \ell$ and $t c \ell \ell$ vector operators with $V_{RR}=1$ and $\Lambda = 1$ TeV. This is overlaid with the SM stacked $VV$, $t\bar t$ and $Z$+jets background processes.
}
\label{fig:inclusive_mll}
\end{figure}

In Tables \ref{tab:mumu_Yields} and \ref{tab:emumu_Yields} we list the expected number of background 
events per 140 fb$^{-1}$
of integrated luminosity, for the di-muon + $b$-jet signal $pp \to \mu^+ \mu^- + j_b + X$ and for the inclusive tri-lepton signal $pp \to e \mu^+ \mu^- +X$,
with di-muon invariant mass lower cut selections of $m_{\mu^+ \mu^-}^{\tt min}=500,1000,1500,2000$~GeV. 
In Fig.~\ref{fig:inclusive_mll} we show the $m_{\mu^+ \mu^-}$ distribution of the leading $VV$, $t\bar t$ and $Z$+jets SM backgrounds and of the $(e \mu \mu)$ and $(\mu \mu 1b)$ signal scenarios for both the $t u \ell \ell$ and $t c \ell \ell$ $V_{RR}$ operators.

\begin{table*}[htb]

\caption{Number of reducible di-lepton background events per 140 fb$^{-1}$ of integrated luminosity, expected from the SM processes 
$pp \to VV, t \bar t, Z + {\tt jets}, tW \to 
\mu^+ \mu^- + j_b +X$, with $m_{\mu^+ \mu^-}^{\tt min}=500,1000,1500,2000$~GeV.  
See also text. 
\label{tab:mumu_Yields}}
\begin{tabular}{c|c|c|c|c|}
\multicolumn{5}{c|}{Number of background $pp \to \mu \mu + j_b +X$ events/140 fb$^{-1}$} 
\tabularnewline
\cline{1-5} 
sub-process & $m_{\mu^+ \mu^-}^{\tt min}=500$~GeV & $m_{\mu^+ \mu^-}^{\tt min}=1000$~GeV & $m_{\mu^+ \mu^-}^{\tt min}=1500$~GeV & $m_{\mu^+ \mu^-}^{\tt min}=2000$~GeV 
\tabularnewline
\hline \hline
\
\multirow{5}{*}{}
$VV$ & 13.0 & 1.2 & 0.2 & 0.1 \tabularnewline
\
$t \bar t$ & 336.4 & 6.9 & 0.3 & 0.0 \tabularnewline
\
$Z + {\tt jets}$ & 128.2 & 10.9 & 1.9 & 0.4 \tabularnewline
\
$Wt$ & 67.1 & 1.3 & 0.1 & 0.0 \tabularnewline
\hline
\hline
\multicolumn{1}{c|}{Total $\mu \mu j_b$ Background events} & 477.7 & 19.1 & 2.4 & 0.5 \tabularnewline
\hline
\hline
\end{tabular}
\end{table*}

\begin{table*}[htb]
\caption{Same as Table \ref{tab:mumu_Yields} but for the case of the 
$(e \mu \mu)$ signal $p p \to e \mu^+ \mu^- + X$. \label{tab:emumu_Yields}}
\begin{tabular}{c|c|c|c|c|}
\multicolumn{5}{c|}{Number of background $pp \to e \mu \mu +X$ events/140 fb$^{-1}$} 
\tabularnewline
\cline{1-5} 
 sub-process & $m_{\mu^+ \mu^-}^{\tt min}=500$~GeV & $m_{\mu^+ \mu^-}^{\tt min}=1000$~GeV & $m_{\mu^+ \mu^-}^{\tt min}=1500$~GeV & $m_{\mu^+ \mu^-}^{\tt min}=2000$~GeV 
\tabularnewline
\hline \hline
\
\multirow{5}{*}{}
$VV$ & 7.1 & 1.0 & 0.2 & 0.1  \tabularnewline
\
$t \bar t$ & 78.2 & 1.6 & 0.1 & 0.0  \tabularnewline
\
$Z + {\tt jets}$ & 16.5 & 1.3 & 0.2 & 0.0 \tabularnewline
\
$Wt$ & 9.5 & 0.2 & 0.0 & 0.0 \tabularnewline
\hline
\hline
\multicolumn{1}{c|}{Total $e \mu \mu$ Background events} & 111.8 & 4.1 & 0.5 & 0.1 \tabularnewline
\hline
\hline
\end{tabular}
\end{table*}

\subsection{Results: sensitivity and bounds}

For a sensitivity study (i.e. placing a bound on the NP scale $\Lambda$), we calculated the $p$-value for each signal and background hypothesis using the
\verb|BinomialExpP| function by \verb|RooFit|~\cite{Verkerke:2003ir}.
We calculate the $p$-value of the background-only and background+signal hypotheses for each point and then perform a $CL_{s}$~\cite{Read:2002hq} test to determine the 95\% Confidence Level (CL) exclusion values for $\Lambda$.
We then find an optimized selection of a minimum OS di-muon invariant mass cut, $m_{\mu\mu}^{\tt min}$, which yields the best limit on $\Lambda$ in each channel, where at least one expected event was demanded for each one of the signal hypotheses.
\begin{figure}[H]
\centering
\includegraphics[width=0.30\textwidth]{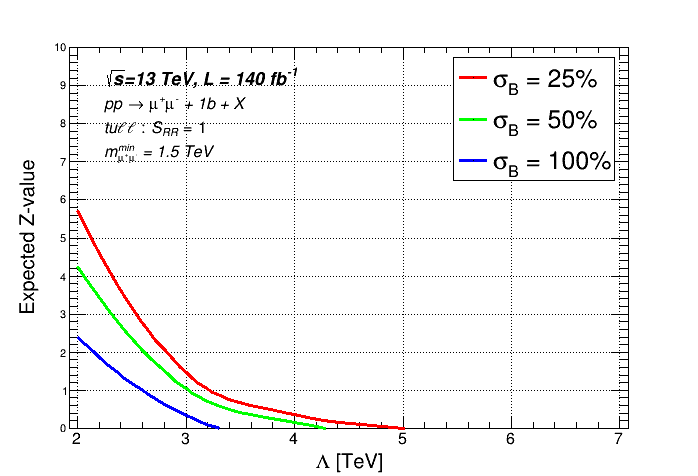}
\includegraphics[width=0.30\textwidth]{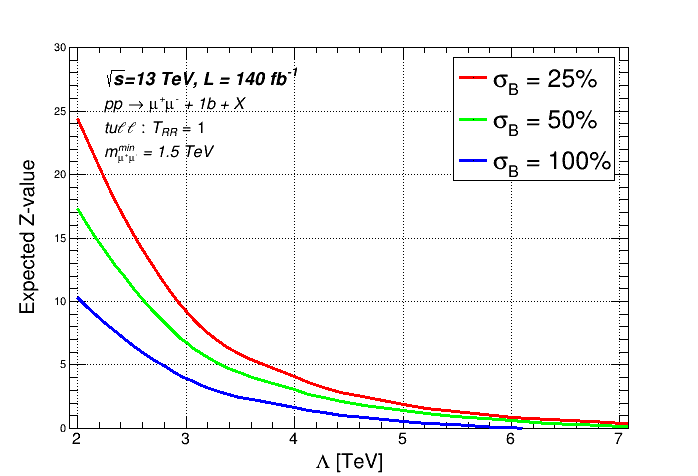}
\includegraphics[width=0.30\textwidth]{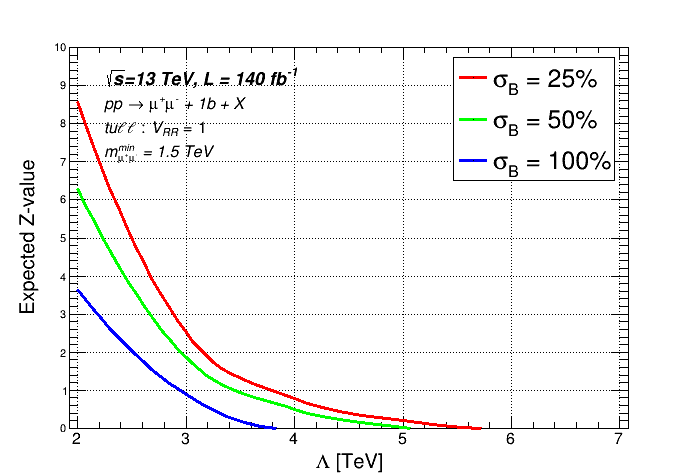}\\
\includegraphics[width=0.30\textwidth]{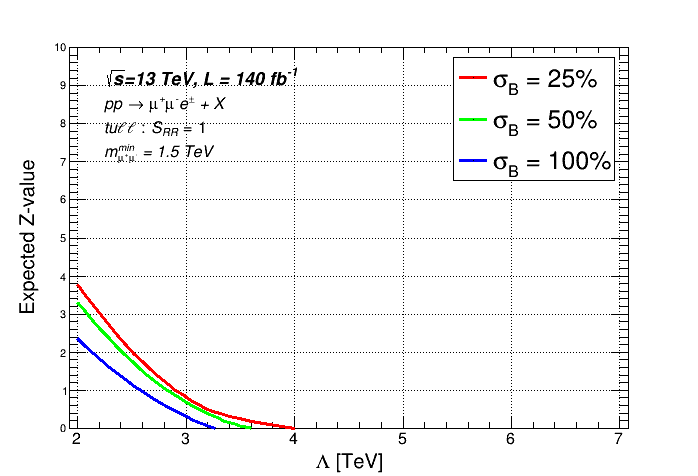}
\includegraphics[width=0.30\textwidth]{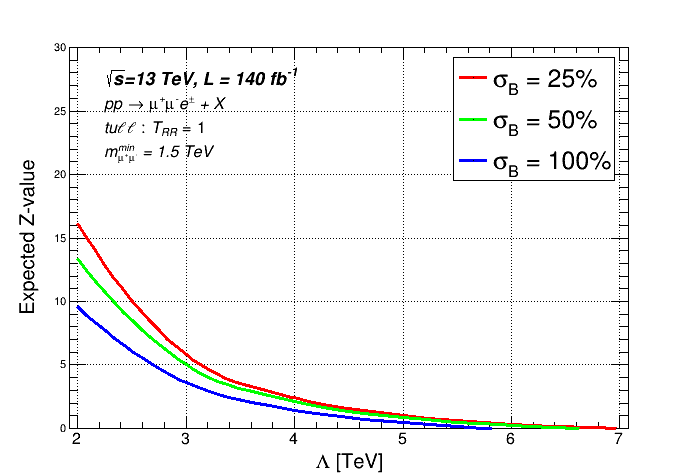}
\includegraphics[width=0.30\textwidth]{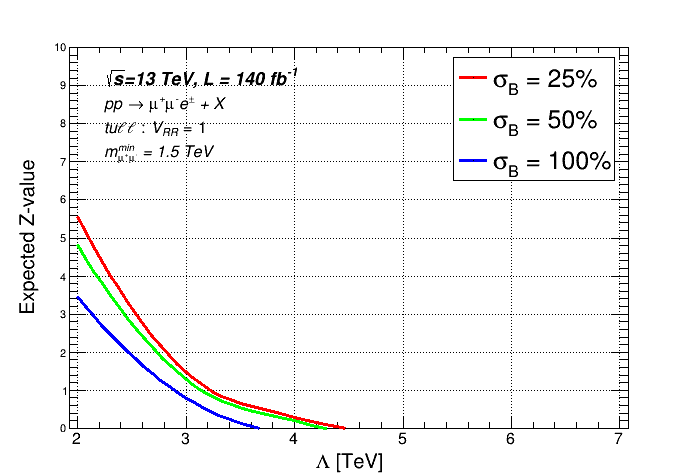}
\caption{Expected $Z$-value for the signal hypotheses varied with respect to the scale $\Lambda$, of the $tu \ell \ell$ scalar, tensor and vector operators with $S_{RR}=1$ (left), $T_{RR}=1$ (middle) and $V_{RR}=1$ (right), for an integrated luminosity of 140~fb$^{-1}$ and with $m_{\mu^+\mu^-}^{\tt min} = 1.5$~TeV. The $(\mu \mu 1b)$ and $(e \mu \mu)$ final state selections, $pp \to \mu^+ \mu^- + j_b + X $ (upper) and $pp \to e \mu^+ \mu^- +X $ (bottom), respectively, are presented.}
\centering
\label{fig:Z_value}
\end{figure}
The expected $Z$-value, which is defined as the number of standard deviations from the background-only hypothesis given a signal yield and background uncertainty, is calculated by the \verb|BinomialExpZ| function by \verb|RooFit|~\cite{Verkerke:2003ir}.
Examples of the expected $Z$-value from the tri-lepton signal are plotted in Fig.~\ref{fig:Z_value}, as a function of $\Lambda$ for the case of the $tu \ell \ell$ operator and for several values of the relative overall uncertainty, $\sigma_B=25\%,50\%,100\%$, with the currently available integrated luminosity of 140~fb$^{-1}$. Clearly, our results depend on the relative uncertainty. Furthermore, the relative uncertainties for the $(e \mu \mu)$ and $(\mu \mu 1b)$ signal selections may be different and that can determine which of the two is more adequate for searching for the 4-Fermi signal scenarios discussed in this paper.Keeping that in mind, we analyze both the $(e \mu \mu)$ and $(\mu \mu 1b)$ signal channels and choose a benchmark value of $\sigma_B = 25\%$ (see e.g., \cite{bbll}) for both of them, assuming that the signal uncertainty is included within $\sigma_B$.\footnote{We note that the statistical uncertainty from the event generator is of ${\cal O}(1\%)$ and is considered within this approximation.}

\begin{figure}[]
\centering
\includegraphics[width=0.45\textwidth]{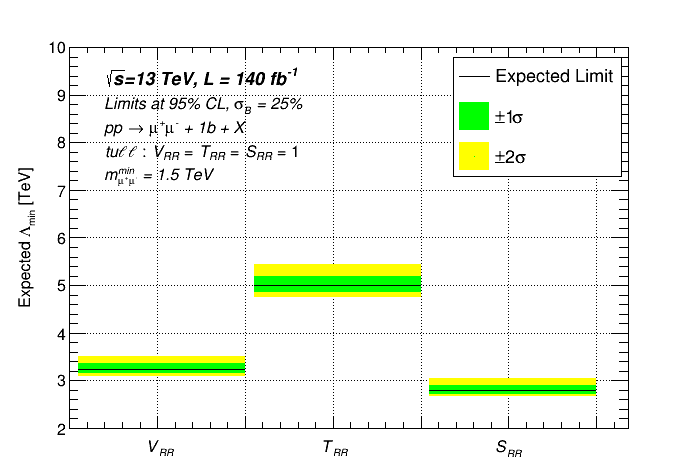}
\includegraphics[width=0.45\textwidth]{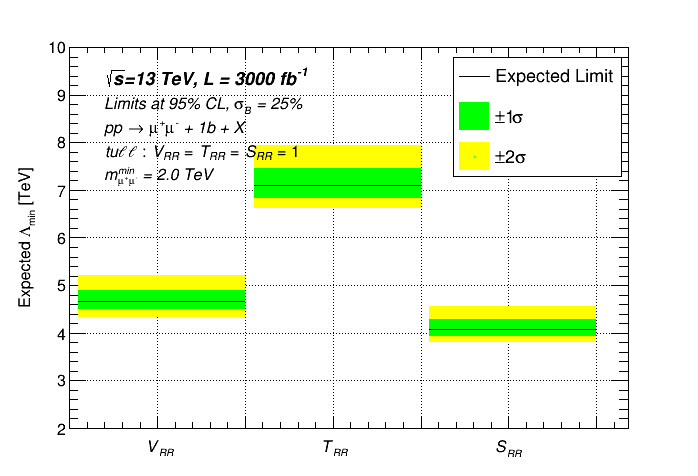} \\
\includegraphics[width=0.45\textwidth]{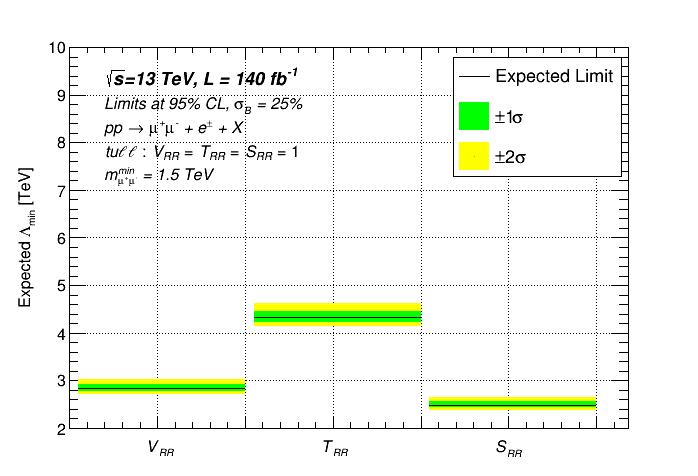}
\includegraphics[width=0.45\textwidth]{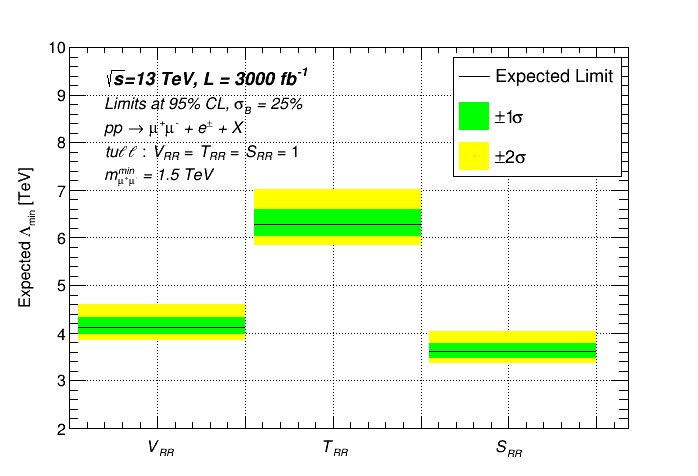}
\caption{
Expected 95\%~CL upper limit on $\Lambda$, $\Lambda_{\tt min}(95\%~CL)$, of the $tu \ell \ell$ operator for 3 signal scenarios: $S_{RR}=1$, $T_{RR}=1$ and $V_{RR}=1$, and total integrated luminosities of 140~fb$^{-1}$ (left) and 3000~fb$^{-1}$ (right). 
The $(\mu \mu 1b)$ and $(e \mu \mu)$ final states selection are presented in the upper and lower plots, respectively.
For all cases $m_{\mu^+ \mu^-}^{\tt max}=5$~TeV and the optimal $m_{\mu^+ \mu^-}^{\tt min}$ selections were used (see also Tables~\ref{tab:mumu_1b_selections:tull}-\ref{tab:mumu_e_selections:tull}).
Also, for all cases the overall uncertainty is chosen to be 25\% at $1\sigma$ as explained in the text.}
\centering
\label{fig:Exclusion}
\end{figure}

\begin{figure}[]
\centering
\includegraphics[width=0.45\textwidth]{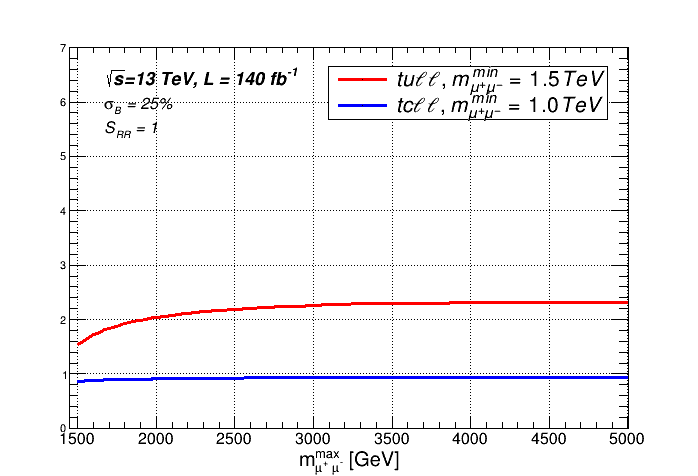}
\includegraphics[width=0.45\textwidth]{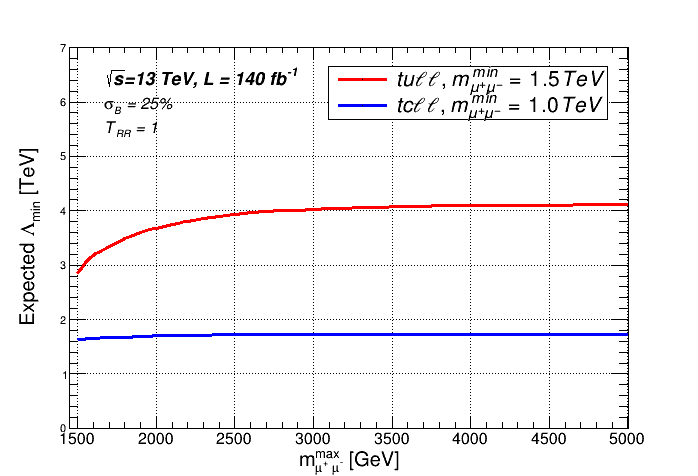}
\caption{
Expected 95\%~CL upper limit on the scale $\Lambda$ of the $tu \ell \ell$ and $tc \ell \ell$ operators for the $S_{RR}$ (left) and $T_{RR}$ signal scenarios, as a function of the upper invariant mass selection, $m_{\mu^+ \mu^-}^{\tt max}$. Results are shown for a total integrated luminosity of 140~fb$^{-1}$. In both cases we assume 25\% background uncertainty at $1\sigma$.}
\centering
\label{fig:Exclusion_tull}
\end{figure}

In Tables~\ref{tab:mumu_1b_selections:tull}-\ref{tab:mumu_e_selections:tull} we present the expected
95\%~CL bounds on the scale $\Lambda$ of the $tull$ and $tcll$ operators, for the 3 different signal scenarios: $S_{RR} = 1, T_{RR} = 1$ and $V_{RR} = 1$. The 95\%~CL bounds on the scale of the 
$tu \ell\ell$ operator are also depicted in Fig.~\ref{fig:Exclusion},
for the optimized $m_{\mu^+ \mu^-}^{\tt min}$ selection which yields the best expected limit for each case, along with the 
$\pm 1 \sigma$  and $\pm 2 \sigma$ bands, as explained below.
As mentioned above, the two integrated luminosity scenarios ${\cal L}= 140$ and $3000$ fb$^{-1}$ are considered. An upper cut of 
$m_{\mu^+ \mu^-}^{\tt max}=5$~TeV and $m_{\mu^+ \mu^-}^{\tt max}=3$~TeV were applied on the OSSF di-muons in the $tu \ell\ell$ and 
$tc \ell\ell$ cases, respectively, in order to be within the EFT validity regime, as discussed above. We note, though, that the 95\%~CL bounds reported here are rather mildly sensitive 
to $m_{\mu^+ \mu^-}^{\tt max}$, as illustrated in Fig.~\ref{fig:Exclusion_tull} for the $tu \ell \ell$ and $tc \ell \ell$ scalar ($S_{RR}$) and tensor operators ($T_{RR}$). 
\begin{table*}[]
  \caption{Expected maximum 95\%~CL sensitivity ranges to the scale $\Lambda$, $\Lambda_{\tt min}(95\%~CL)$, of the $t u \ell \ell$ and $t c \ell \ell$ 4-Fermi operators, obtained via the di-lepton $(\mu \mu 1b)$ signal $pp \to \mu^+ \mu^- + j_b +X$ with the corresponding optimal $m_{\mu^+ \mu^-}^{\tt min}$ selection. An upper selection on the di-muon invariant mass of $m_{\mu^+ \mu^-}^{\tt max} = 3, ~5$~TeV was applied in the $t c \ell \ell$, $t u \ell \ell$ cases, respectively.
  Results are shown for the 3 signal scenarios of each operator: $S_{RR} = 1, T_{RR} = 1, V_{RR} = 1$. See also text and caption of Fig.~\ref{fig:Exclusion}.}
  \label{tab:mumu_1b_selections:tull}
\begin{tabular}{c|c|c|c|c|c|}
\multirow{1}{*}{} \
  & \multicolumn{5}{c|}{Expected bounds $\Lambda_{\tt min}(95\%~CL)$ [TeV]; $(\mu \mu 1b): pp \to \mu^+ \mu^- +j_b +X$}
\tabularnewline
\cline{2-6} 
  & Operator & \multicolumn{2}{c|}{\boldmath{$tu \mu\mu$} {\bf 4-Fermi case}} & \multicolumn{2}{c|}{
  \boldmath{$tc \mu\mu$} {\bf 4-Fermi case}}
  \tabularnewline
\cline{2-6} 
  & Coupling & $m_{\mu^+ \mu^-}^{\tt min}$~[GeV] & $\Lambda_{\tt min}(95\%~CL)$ [TeV] & $m_{\mu^+ \mu^-}^{\tt min}$~[GeV] & $\Lambda_{\tt min}(95\%~CL)$ [TeV] 
  \tabularnewline
\hline \hline
\
 & $S_{RR}=1$   & & $2.8^{+0.1}_{-0.1}$ & & $1.0^{+0.1}_{-0.1}$ \noeft \tabularnewline
\
${\cal L}=140$ fb$^{-1}$ & $T_{RR}=1$ 
 & $1500$ & $5.0^{+0.1}_{-0.2}$ & $1000$ & $1.8^{+0.1}_{-0.1}$ \phantom{\noeft} \tabularnewline
\
& $V_{RR}=1$
  & & $3.2^{+0.1}_{-0.1}$ & & $1.1^{+0.1}_{-0.1}$ \phantom{\noeft}  \tabularnewline
\hline 
\hline 
 & $S_{RR}=1$
   &  & $4.1^{+0.1}_{-0.2}$ & & $1.3^{+0.1}_{-0.1}$ \noeft \tabularnewline
\
${\cal L}=3000$ fb$^{-1}$ & $T_{RR}=1$
  & $2000$ & $7.1^{+0.3}_{-0.3}$ & $1500$ & $2.4^{+0.1}_{-0.1}$\phantom{\noeft} \tabularnewline
\
& $V_{RR}=1$
  &  & $4.7^{+0.2}_{-0.2}$ & & $1.5^{+0.1}_{-0.1}$ \noeft \tabularnewline
\hline
\hline
\end{tabular}
\end{table*}
\begin{table*}[]
  \caption{Same as Table \ref{tab:mumu_1b_selections:tull} but for the 
$(e \mu \mu)$ signal $p p \to e \mu^+ \mu^- + X$.} 
  \label{tab:mumu_e_selections:tull}
\begin{tabular}{c|c|c|c|c|c|}
\multirow{1}{*}{} \
  & \multicolumn{5}{c|}{Expected bounds $\Lambda_{\tt min}(95\%~CL)$ [TeV]; $(e\mu \mu): pp \to e \mu^+ \mu^- + X$}
\tabularnewline
\cline{2-6} 
  & Operator & \multicolumn{2}{c|}{\boldmath{$tu \mu\mu$} {\bf 4-Fermi case}} & \multicolumn{2}{c|}{
  \boldmath{$tc \mu\mu$} {\bf 4-Fermi case}}
  \tabularnewline
\cline{2-6} 
  & Coupling & $m_{\mu^+ \mu^-}^{\tt min}$~[GeV] & $\Lambda_{\tt min}(95\%~CL)$ [TeV] & $m_{\mu^+ \mu^-}^{\tt min}$~[GeV] & $\Lambda_{\tt min}(95\%~CL)$ [TeV] 
  \tabularnewline
\hline \hline
\
 & $S_{RR}=1$   & & $2.3^{+0.0}_{-0.1}$ & & $0.9^{+0.0}_{-0.0}$ \noeft \tabularnewline
\
${\cal L}=140$ fb$^{-1}$ & $T_{RR}=1$ 
 & $1500$ & $4.1^{+0.1}_{-0.1}$ & $1000$ & $1.7^{+0.1}_{-0.1}$ \phantom{\noeft} \tabularnewline
\
& $V_{RR}=1$
  & & $2.7^{+0.0}_{-0.1}$ & & $1.1^{+0.0}_{-0.0}$\phantom{\noeft}  \tabularnewline
\hline 
\hline 
 & $S_{RR}=1$
   &  & $3.5^{+0.1}_{-0.1}$ & & $1.1^{+0.1}_{-0.1}$\phantom{\noeft}  \tabularnewline
\
${\cal L}=3000$ fb$^{-1}$ & $T_{RR}=1$
  & $1500$ & $6.3^{+0.2}_{-0.3}$ & $1000$ & $2.1^{+0.1}_{-0.1}$ \phantom{\noeft} \tabularnewline
\
& $V_{RR}=1$
  &  & $4.1^{+0.1}_{-0.2}$ & & $1.3^{+0.1}_{-0.1}$\phantom{\noeft}  \tabularnewline
\hline
\hline
\end{tabular}
\end{table*}

\newpage

\begin{table*}[htb]
  \caption{Expected discovery potential for the scale of NP $\Lambda$,  of the $t u \ell \ell$ and $t c \ell \ell$ 4-Fermi operators, obtained via the di-lepton $(\mu \mu 1b)$ signal $pp \to \mu^+ \mu^- + j_b +X$ with the corresponding optimal $m_{\mu^+ \mu^-}^{\tt min}$ selection. An upper selection on the di-muon invariant mass of $m_{\mu^+ \mu^-}^{\tt max} = 3, ~5$~TeV was applied in the $t c \ell \ell$, $t u \ell \ell$ cases, respectively.
  Results are shown for the 3 signal scenarios of each operator: $S_{RR} = 1, T_{RR} = 1, V_{RR} = 1$.}
  \label{tab:mumu_1b_discovery:tull}
\begin{tabular}{c|c|c|c|c|c|}
\multirow{1}{*}{} \
  & \multicolumn{5}{c|}{Expected discovery potential $\Lambda(5 \sigma)$ [TeV]; $(\mu \mu 1b): pp \to \mu^+ \mu^- +j_b +X$}
\tabularnewline
\cline{2-6} 
  & Operator & \multicolumn{2}{c|}{\boldmath{$tu \mu\mu$} {\bf 4-Fermi case}} & \multicolumn{2}{c|}{
  \boldmath{$tc \mu\mu$} {\bf 4-Fermi case}}
  \tabularnewline
\cline{2-6} 
  & Coupling & $m_{\mu^+ \mu^-}^{\tt min}$~[GeV] & $\Lambda (5\sigma)$ [TeV] & $m_{\mu^+ \mu^-}^{\tt min}$~[GeV] & $\Lambda (5\sigma)$ [TeV] 
  \tabularnewline
\hline \hline
\
 & $S_{RR}=1$   & & $2.1$ & & $0.7$ \noeft \tabularnewline
\
${\cal L}=140$ fb$^{-1}$ & $T_{RR}=1$ 
 & $1500$ & $3.7$ & $1000$ & $1.4$ \phantom{\noeft} \tabularnewline
\
& $V_{RR}=1$
  & & $2.4$ & & $0.9$ \noeft \tabularnewline
\hline 
\hline 
 & $S_{RR}=1$
   &  & $3.1$ & & $1.0$ \noeft \tabularnewline
\
${\cal L}=3000$ fb$^{-1}$ & $T_{RR}=1$
  & $2000$ & $5.3$ & $1500$ & $1.8$ \phantom{\noeft} \tabularnewline
\
& $V_{RR}=1$
  &  & $3.5$ & & $1.1$ \noeft \tabularnewline
\hline
\hline
\end{tabular}
\end{table*}
\begin{table*}[htb]
  \caption{Same as Table \ref{tab:mumu_1b_discovery:tull} but for the 
$(e \mu \mu)$ signal $p p \to e \mu^+ \mu^- + X$.}
  \label{tab:mumu_e_discovery:tull}
\begin{tabular}{c|c|c|c|c|c|}
\multirow{1}{*}{} \
  & \multicolumn{5}{c|}{Expected discovery potential $\Lambda(5 \sigma)$ [TeV]; $(e \mu \mu): pp \to e \mu^+ \mu^- + X$}
\tabularnewline
\cline{2-6} 
  & Operator & \multicolumn{2}{c|}{\boldmath{$tu \mu\mu$} {\bf 4-Fermi case}} & \multicolumn{2}{c|}{
  \boldmath{$tc \mu\mu$} {\bf 4-Fermi case}}
  \tabularnewline
\cline{2-6} 
  & Coupling & $m_{\mu^+ \mu^-}^{\tt min}$~[GeV] & $\Lambda (5\sigma)$ [TeV] & $m_{\mu^+ \mu^-}^{\tt min}$~[GeV] & $\Lambda (5\sigma)$ [TeV] 
  \tabularnewline
\hline \hline
\
 & $S_{RR}=1$   & & $1.7$ & & $0.7$ \noeft \tabularnewline
\
${\cal L}=140$ fb$^{-1}$ & $T_{RR}=1$ 
 & $1500$ & $3.0$ & $1000$ & $1.3$ \phantom{\noeft} \tabularnewline
\
& $V_{RR}=1$
  & & $1.9$ & & $0.8$ \noeft \tabularnewline
\hline 
\hline 
 & $S_{RR}=1$
   &  & $2.7$ & & $0.9$ \noeft \tabularnewline
\
${\cal L}=3000$ fb$^{-1}$ & $T_{RR}=1$
  & $1500$ & $4.7$ & $1000$ & $1.6$ \phantom{\noeft} \tabularnewline
\
& $V_{RR}=1$
  &  & $3.0$ & & $1.0$ \noeft \tabularnewline
\hline
\hline
\end{tabular}
\end{table*}

\subsection{Results: discovery potential}

An estimate of the discovery potential can be inferred from the expected $Z$-values mentioned above;
in particular, $Z=5$ corresponds to a $5\sigma$ discovery.
Once again, as a benchmark selection, we assume that the relative overall uncertainty is 25\% and in  
Tables~\ref{tab:mumu_1b_discovery:tull}-\ref{tab:mumu_e_discovery:tull} we list the 
expected $5 \sigma$ discovery potential, $\Lambda(5 \sigma)$, for the 3 different signal scenarios: $S_{RR} = 1, T_{RR} = 1, V_{RR} = 1$ of both the $t u\ell \ell$ and $tc \ell \ell$ 4-Fermi operators, via the $(\mu \mu 1b)$ and $(e \mu \mu)$ signal selections, respectively. Results are again shown for the two integrated luminosity cases corresponding to the currently accumulated LHC data and the planned HL-LHC luminosity. We see e.g., that a $5 \sigma$ discovery of the $tu \ell \ell$ tensor interactions is possible within the current LHC accumulated data if $\Lambda \lsim 3.7$~TeV with the $(\mu \mu 1b)$ selection and $\Lambda \lsim 3$~TeV in the tri-lepton $(e \mu \mu)$ case.

\newpage

\subsection{Results: sensitivity at a future 27 and 100~TeV hadron colliders}

\begin{figure}[htb]
\centering
\includegraphics[width=0.45\textwidth]{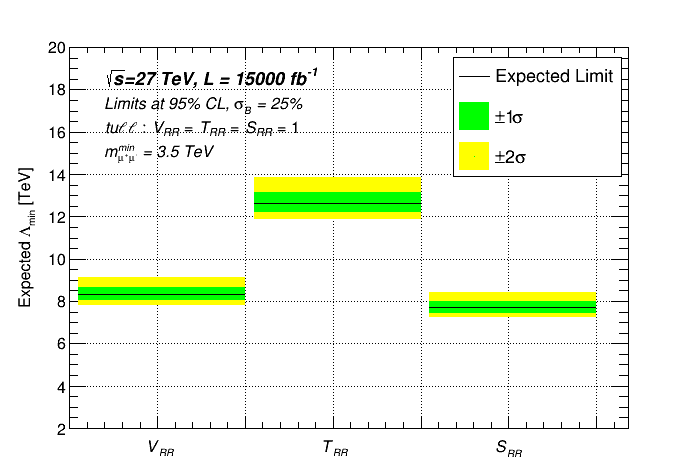}
\includegraphics[width=0.45\textwidth]{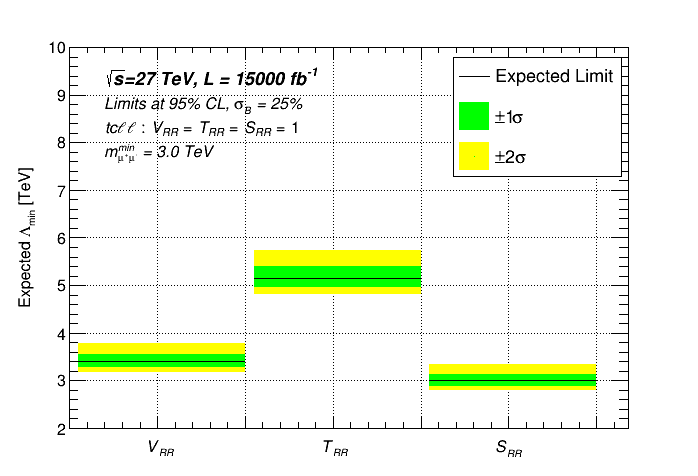}\\
\includegraphics[width=0.45\textwidth]{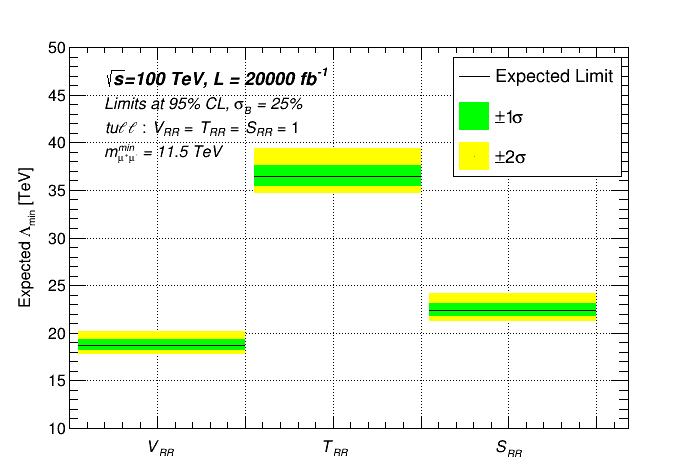}
\includegraphics[width=0.45\textwidth]{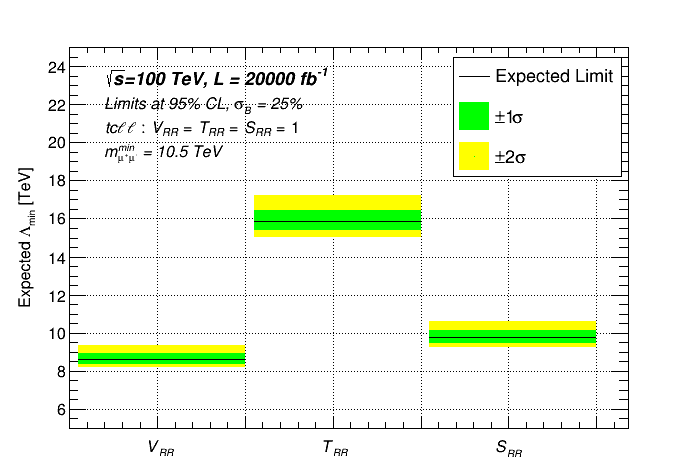}
\caption{
Expected 95\%~CL upper limit on $\Lambda$ of the $tu \ell \ell$ (left) and $tc \ell \ell$ (right) operators for centre-of-mass energy of 27~TeV (upper plots) and 100~TeV (lower plots), for 3 signal scenarios: $S_{RR}=1$, $T_{RR}=1$ and $V_{RR}=1$. The total integrated luminosity is 15000~fb$^{-1}$ for the 27~TeV machine and 20000~fb$^{-1}$ for 100~TeV one.
Results are shown with 25\% overall relative uncertainty at $1\sigma$ and 
with $m_{\mu^+ \mu^-}^{\tt min}$ selections as indicated. See also text.}
\centering
\label{fig:Exclusion_27TeV}
\end{figure}

We have also extended our study to future hadron machines; specifically, to the sensitivity of a 27~TeV and a 100~TeV proton-proton collider to the $tu_i \ell \ell$ operators, for which we have assumed a total integrated luminosity of 15000 fb$^{-1}$~\cite{Azzi:2019yne} and 20000 fb$^{-1}$~\cite{Mangano:2016jyj}, respectively. The expected 95\% CL bounds 
for the higher energy proton-proton colliders are presented in Fig.~\ref{fig:Exclusion_27TeV}, where
we show the 95\%~CL upper bounds on the scale of the $tu \ell \ell$ and $tc \ell \ell$ 4-Fermi operators for the 3 signal scenarios $S_{RR}=1$, $T_{RR}=1$ and $V_{RR}=1$. Here also, 
the expected bounds are presented for the optimized $m_{\mu^+ \mu^-}^{\tt min}$ selections and an upper cut on the OSSF muons of $m_{\mu^+ \mu^-}^{\tt max} = 10,30$~TeV for the 27~TeV and 100~TeV cases, respectively,  
and with a $25\%(1\sigma)$ relative uncertainty.

We see from Fig.~\ref{fig:Exclusion_27TeV} that a 27~TeV (100~TeV) proton collider is expected to be sensitive (at 95\%~CL) 
to $\Lambda \gsim 8-13$~TeV($\Lambda \gsim 19-37$~TeV) for the $tu \ell \ell$ operator 
and to $\Lambda \gsim 3-5$~TeV($\Lambda \gsim 9-16$~TeV) for the $tc \ell \ell$ one. This should be compared with the expected reach at other future colliders, such as the proposed Circular Electron Positron Collider (CEPC, see e.g., \cite{CEPCStudyGroup:2018ghi})  and Compact Linear Collider (CLIC, see e.g., \cite{1812.02093} and references therein) $e^+e^-$-machines and the $ep$ Large Hadron-Electron Collider (LHeC, see e.g., \cite{AbelleiraFernandez:2012cc}). For example, it was shown in \cite{1906.04573} and \cite{1906.04884} that the CEPC and CLIC machines, respectively, will be sensitive to scales of the $tuee$ and $tcee$ scalar, vector and tensor operators in the range $\Lambda \sim 5-10$ TeV, depending on the center of mass energy of these future $ee$ machines. For the $tuee$ 4-Fermi terms this is comparable to the reach expected at the HL-LHC via our di- and tri-lepton signals (as shown above), whereas for the $tcee$ operators this is about an order of magnitude better than what we found for the HL-LHC setup, while it is comparable to the expected sensitivity at the HE-LHC (i.e., a 27 or 100 TeV LHC upgrade, see Fig.~\ref{fig:Exclusion_27TeV}). On the other hand, the sensitivity of the LHeC machine to the $tu ee$ and $tc ee$ operators via the single-top $ep \to et$ production channel 
that was studied in \cite{1906.04884}, is comparable to what we find for the current LHC data with 140 fb$^{-1}$ of data.  

We note, though, that these future $ee$ and $ep$ machines are sensitive only to the $tu_i ee$ 4-Fermi operators, as opposed to the LHC which, as we show, can probe also 
the di-muon $t u_i \mu \mu$ operators via our di- and tri-lepton signals.

\subsection{Results: final remarks}

To conclude this section, let us recapitulate some of the 
salient features of our findings and also further comment on the potential richness of the multi-lepton signals considered above:
\begin{itemize}
\item The three lepton final states with an additional light and/or $b$-jet can be a useful probe for searching NP in general; their applicability is not restricted to FC process or to a SMEFT parameterization. These final states are rich in observables sensitive to deviations form the SM; some interesting examples are a Forward-Backward asymmetry, energy asymmetry and triple correlation asymmetries, which can be readily constructed from the available energies and the 4-momenta of the charged leptons along with the 4-momenta of the light and/or $b$-jet in the tri-lepton final state \cite{ourreview}. 
\item For the SMEFT parameterization of FC effects, both the di- and tri-lepton signals are significantly more sensitive to the $tu \ell \ell$ than to the $tc \ell \ell$ 4-Fermi interaction  as a result of the larger $u$-quark content in the colliding protons and the importance of the $ug$ fusion diagrams in Fig.~\ref{fig:Feynman}. 
\item An extra selection of exactly one $b$-tagged jet on the di-lepton signature can yield a significantly better sensitivity to the scale of the underlying FC NP (see tables \ref{tab:mumu_1b_selections:tull}-\ref{tab:mumu_e_discovery:tull}).
\item Since there are no significant SM$\times$NP interference effects for the FC EFT we consider, the sensitivity and reach for the other 4-Fermi vector currents, $V_{LL,},V_{RL}$ and $V_{LR}$ will be identical to that of $V_{RR}$ (which is the one studied above).
\end{itemize}
\begin{figure}[htb]
\includegraphics[width=0.45\textwidth]{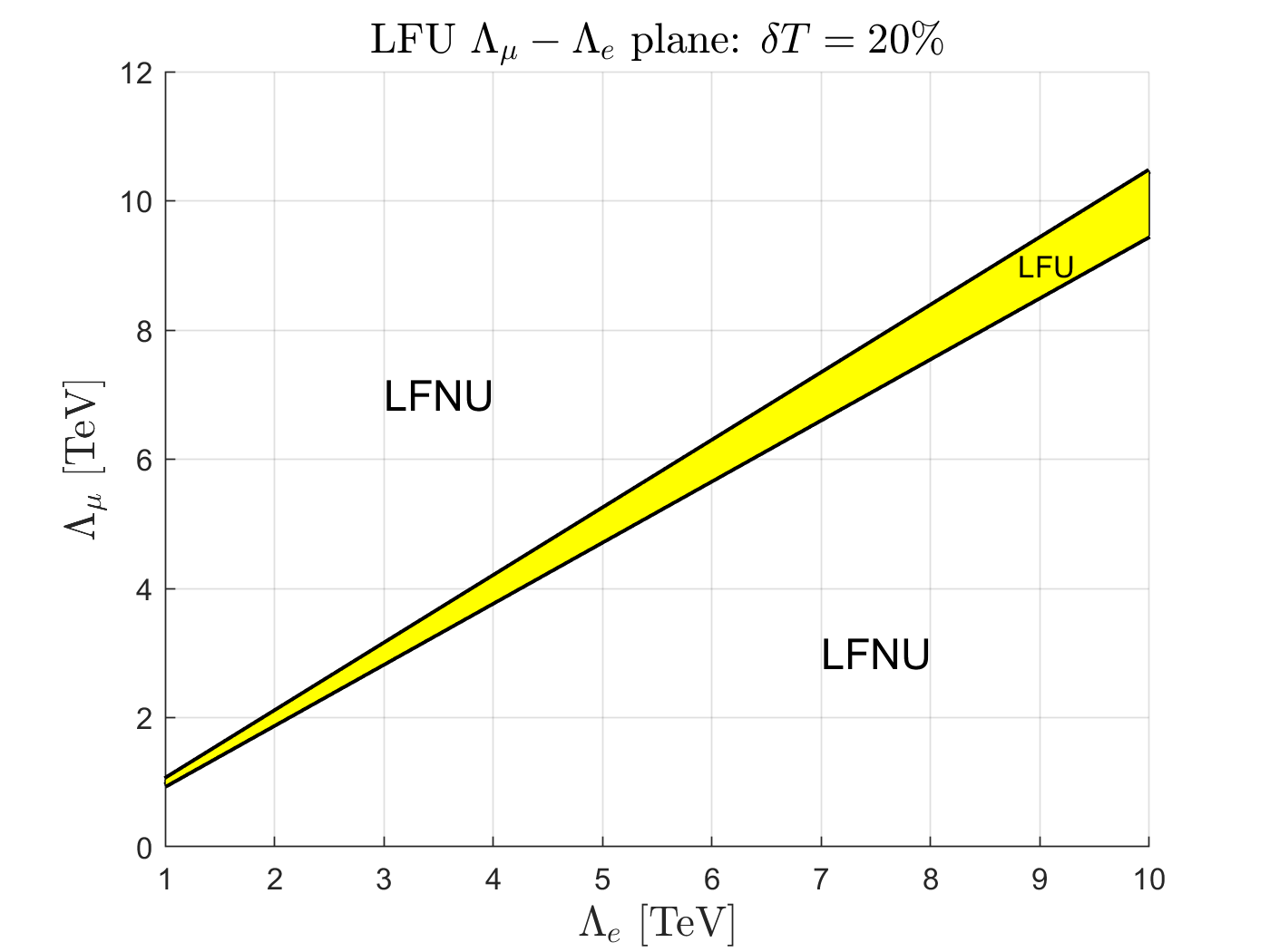}
\includegraphics[width=0.45\textwidth]{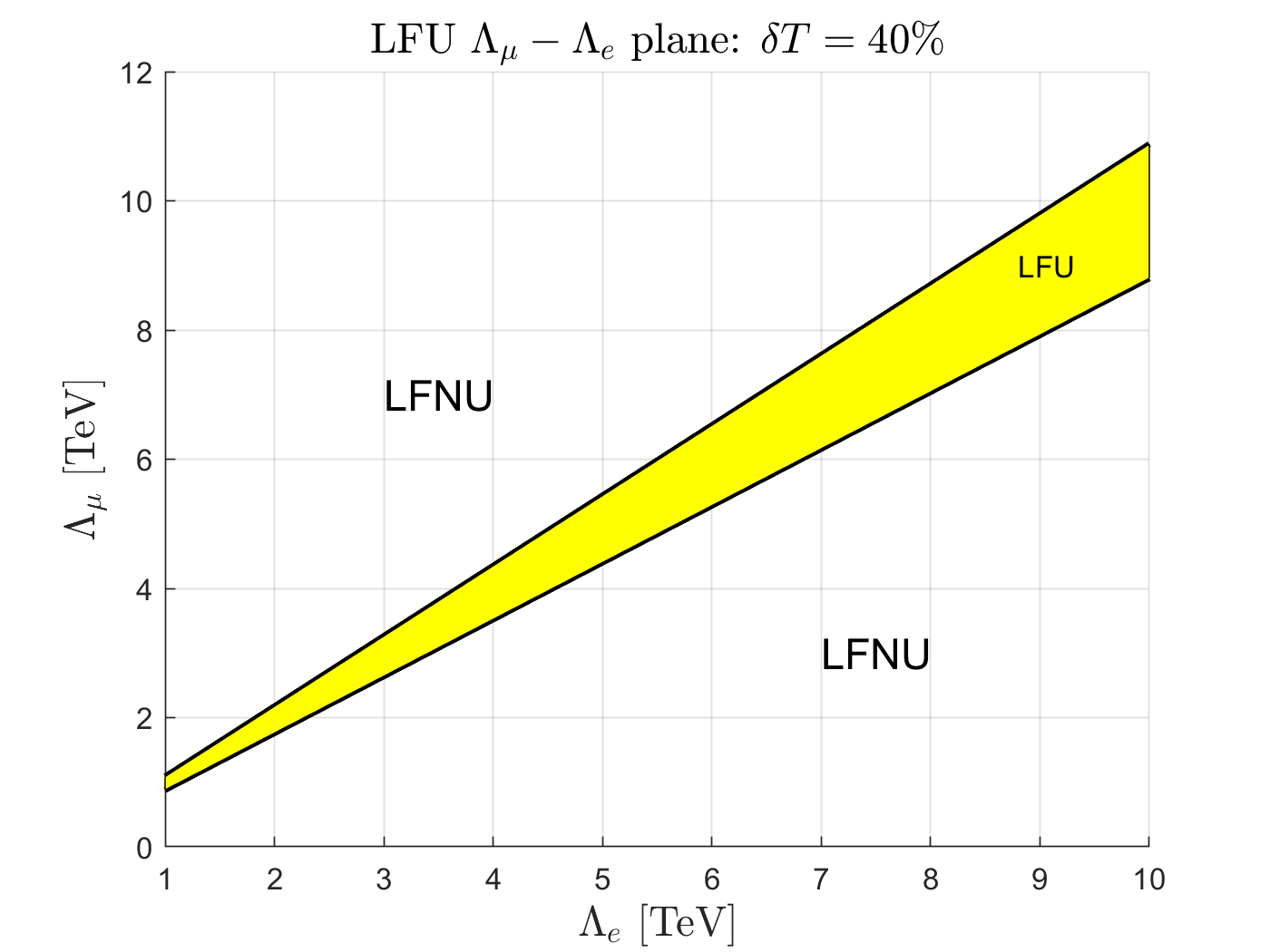}
\includegraphics[width=0.45\textwidth]{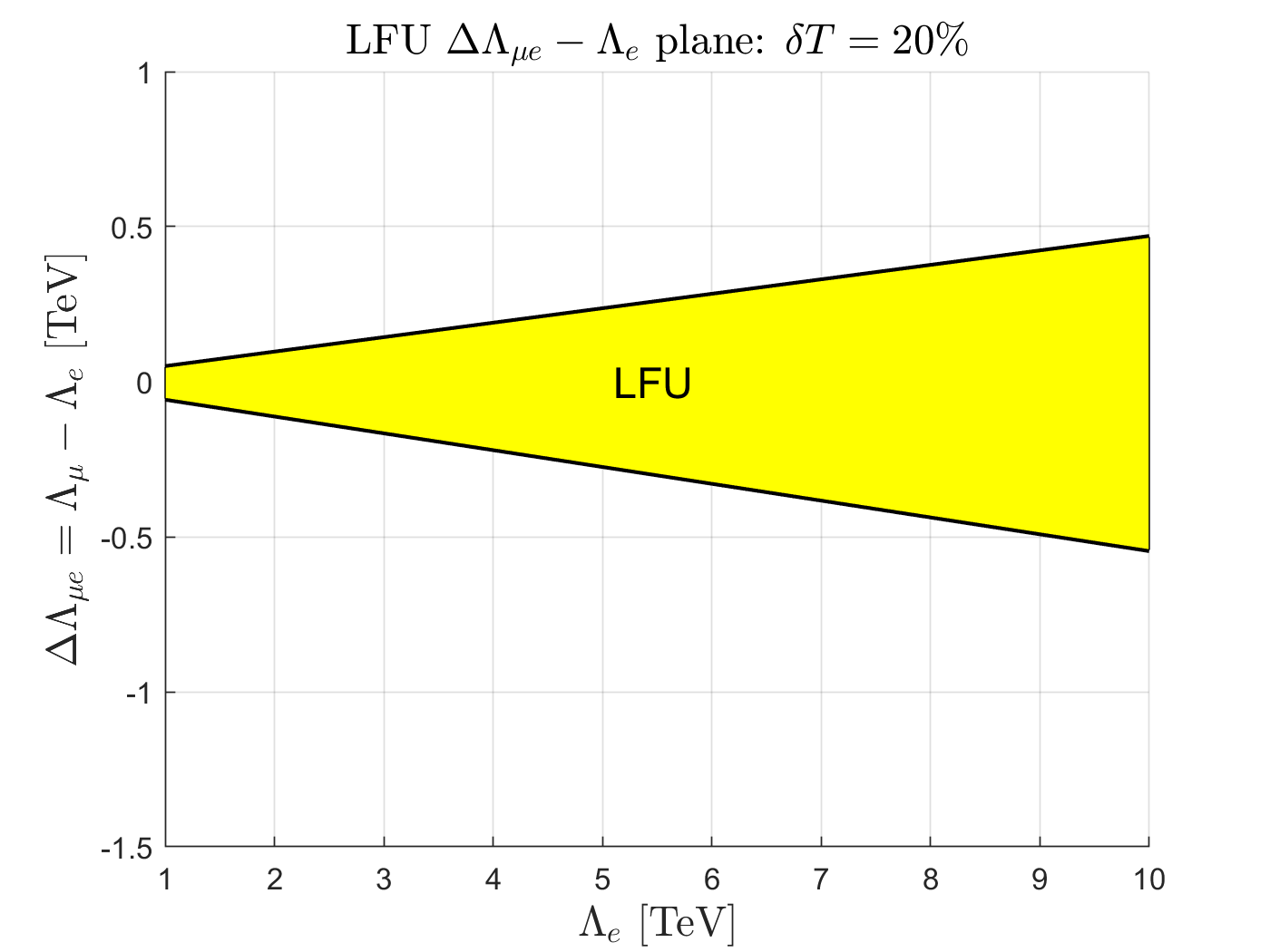}
\includegraphics[width=0.45\textwidth]{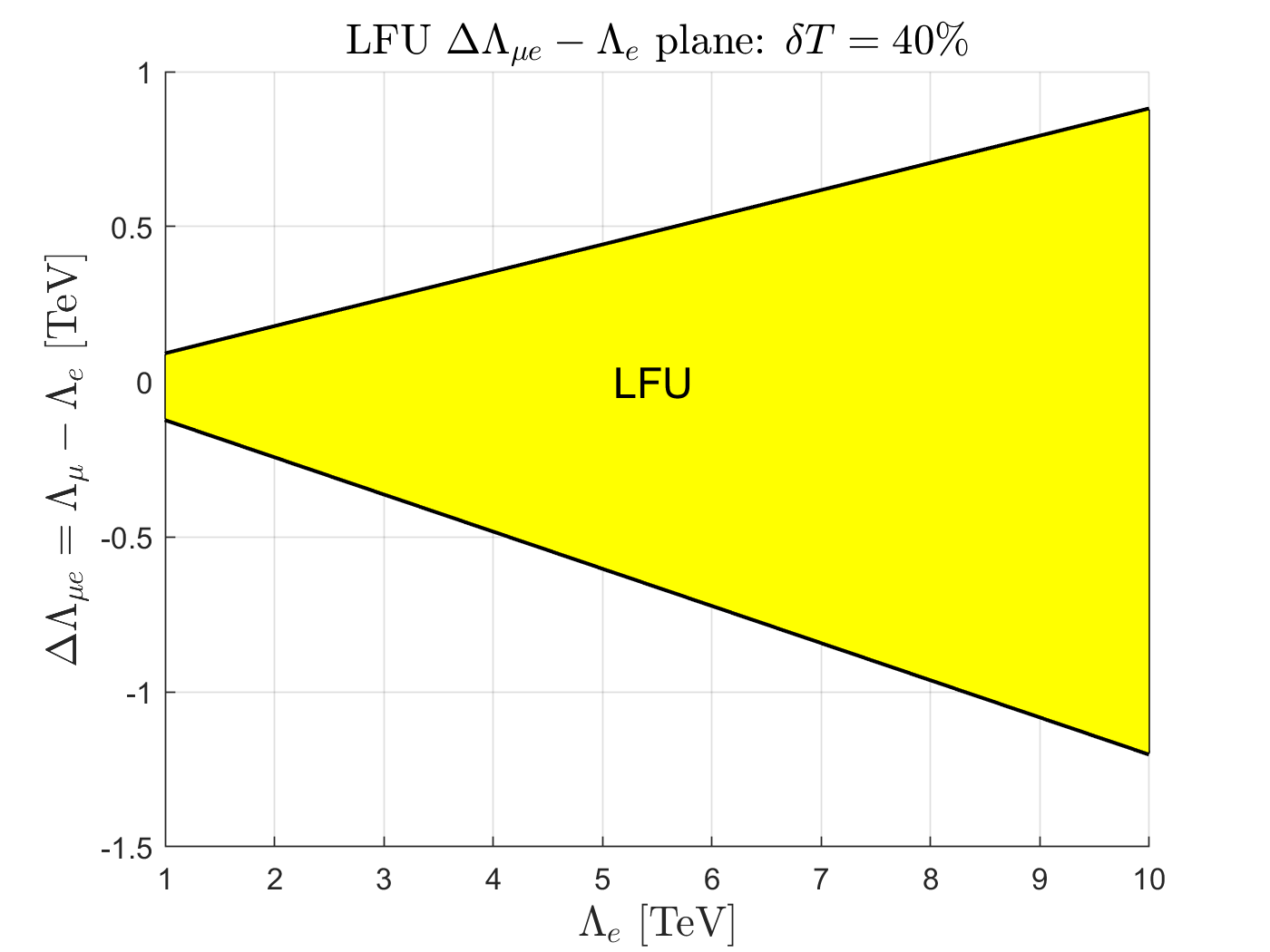}
\caption{Lepton Flavor universal (LFU) regions (yellow shaded) in the $\Lambda_\mu - \Lambda_e$ and $\Lambda_{\mu e} - \Lambda_e$ planes, where $\Lambda_{\mu e} \equiv \Lambda_\mu - \Lambda_e$. The areas outside the yellow shaded areas are where LFU is violated. Two cases are shown: a 20\% (left plots) and 40\% (right plots) uncertainty in the measurement of the LFU T-test in \eqref{Tests_mmm}. See also text.}
\label{fig:LFU}
\end{figure}

\section{Lepton Flavor non-universality}
As mentioned earlier, the di- and tri-lepton signals can also be used to study possible LFNU effects in the FC $tu_i \ell \ell$ 4-Fermi operators  from lepton non-universal effects in the underlying heavy theory, e.g., LFNU couplings of a heavy vector or heavy scalar to the SM leptons.\footnote{The LFNU effects we consider correspond to differences in the $t u ee$, $t u \mu\mu$ and $ tu \tau\tau  $ couplings, and not possible $ t u e \mu $ interactions.} 
Following the work in~\cite{our_LFU_paper}, we can define generic LFNU `tests' for our single-top + di-lepton signals \eqref{eq:llt}, normalized  to the di-electron channels, as follows:
\begin{eqnarray}
T_{{t \ell \ell}_0} =
\frac{\sigma_{(t \ell \ell)_0}}{\sigma_{(tee)_0}}
~~~,~~~ 
T_{{t \ell \ell}_1} =
\frac{\sigma_{(t \ell \ell )_1}}{\sigma_{(tee)_1}}
~; \label{eq:Ttests}
\end{eqnarray}
or, more generally, for our inclusive di- and tri-lepton signals:
\begin{eqnarray}
T_{\ell \ell}^{1b} =
\frac{N(\ell \ell 1b)}{N(e e 1b)}
~~~,~~~ T_{\ell \ell}^{\ell^\prime/\ell^{\prime \prime}} =
\frac{N(\ell^\prime \ell \ell)}{N(\ell^{\prime \prime} e e)} ~, \label{Tests_lll}
\end{eqnarray}
where $N(\ell \ell 1b)$ and $N(\ell^\prime \ell \ell)$ are the number of $pp \to  \ell^+ \ell^- +j_b +X$ and 
$pp \to \ell^\prime \ell^+ \ell^- +X$ events, respectively. An example of an interesting test of LFNU signals is provided by $T_{\mu \mu}^{\mu/e}$  that measures a possible difference in the $tu_i \mu \mu$ and $t u_i e e$ contact interactions. With a selection $m_{\ell \ell}^{\tt min} > 1000$~GeV (ensuring, as shown above, negligible SM background to the NP tri-lepton signals), we have: 
\begin{eqnarray}
T_{\mu \mu}^{\mu/e} =
\frac{N(\mu \mu \mu)}{N(e e e)} \approx \frac{\Lambda_\mu^4}{\Lambda_e^4} ~, \label{Tests_mmm}
\end{eqnarray}
where here $\Lambda_\ell$ denote the scale of the $tu \ell \ell$ operator. 

In Fig.~\ref{fig:LFU} we plot the regions in the $\Lambda_\mu - \Lambda_e$ plane where LFU can be tested with $T_{\mu \mu}^{\mu/e}$, depending on the uncertainty ($\delta T$) in its measurement. We also show in Fig.~\ref{fig:LFU} the size of $\Delta\Lambda_{\mu e} = \Lambda_\mu - \Lambda_e$ splitting required for observing LFNU effects. We see e.g., that $|\Delta\Lambda_{\mu e}| \gsim 0.2$~TeV will yield a measurable LFNU signal if $T_{\mu \mu}^{\mu/e}$ can be measured to a $20\%$ precision.\footnote{Note that ratio observables such as in \eqref{Tests_lll} and \eqref{Tests_mmm} provide more reliable probes of NP, since they potentially minimize the effects of the theoretical uncertainties involved in the calculation of the corresponding cross-sections~\cite{our_LFU_paper}.}

\section{Summary}

We have considered the effects of 4-Fermi $t u_i \ell \ell$ flavor changing interactions ($u_i \in u,c$) in the top-quark sector, which can be generated from different types of underlying heavy physics containing e.g., heavy scalars and/or vectors. We showed that these higher-dimensional FCNC top interactions can lead to new single-top + dilepton signals at the LHC via $pp \to t \ell^+ \ell^-$ and $ pp \to t \ell^+ \ell^- + j$ ($j=$light jet), which can be efficiently probed via the di-lepton + $b$-jet $pp \to \ell^+ \ell^- + j_b +X$ signal and/or in tri-lepton $pp \to \ell^\prime \ell^+ \ell^- + X$ events, containing opposite-sign same-flavor (OSSF) di-leptons, e.g.,  $pp \to e \mu^+ \mu^- + X$, if the NP involves the $t u_i \mu \mu$ contact terms and the top decays via $t \to bW \to b e \nu_e$. 

We have studied in some detail the SM background to these di- and tri-lepton signatures, which is dominated by $pp \to t \bar t, Z+{\tt jets}, WZ$ and showed that an excellent separation between the NP signals and the background can be obtained with a selection of events with high OSSF di-leptons invariant mass $m_{\ell^+ \ell^-}^{\tt min}(OSSF)> 1$ TeV.  The high invariant mass selection on the OSSF di-leptons also allows to isolate the FC $t u_i \ell \ell$ 4-Fermi dynamics from other types of NP, e.g., anomalous FC $t u_i Z$ terms, that may also contribute to the same di- and tri-lepton signals, but with on-$Z$ peak OSSF di-leptons. We also find that an additional selection of a single $b$-tagged jet is useful for tracking the top-quark decay in these events and, in the di-lepton signal case $pp \to \ell^+ \ell^- + j_b +X$, it significantly improves the sensitivity to the scale of these FC $t u_i \ell \ell$ operators. 

We have shown that the current ${\cal O}(1)$ TeV bounds on the scale of these $t u \ell \ell$ and $t c \ell \ell$ FC 4-Fermi interactions (from LEP2 and from $pp \to t \bar t$ followed by $t \to \ell^+ \ell^- j$) can be appreciably improved.
For example, 95\% CL bounds of $\Lambda \lsim 5(3.2) $~TeV are expected on the scale of a tensor(vector) $t u \mu \mu$ interaction, already with the current $\sim 140$ fb$^{-1}$ of LHC 
data, via the di-muon $pp \to \mu^+ \mu^- + j_b +X$ signal; this is an improvement by a factor 
of $3-5$ with respect to the current bounds on these operators. 
The expected reach at the HL-LHC with 3000 fb$^{-1}$ of data is $\Lambda \lsim 7.1(4.7)$~TeV 
for the tensor(vector) FC $t u \ell \ell$ 4-Fermi interactions 
and $\Lambda \lsim 2.4(1.5)$~TeV for the corresponding $t c \mu \mu$ operators.
We have considered the consistency of these bounds with restrictive requirements for the domain of validity of the EFT prescription and imposed the relevant EFT-validity criteria accordingly.  

We have also considered the potential sensitivity of higher energy 27 and 100~TeV proton-proton colliders to the $tu \ell \ell$ and $tc \ell \ell$ 4-Fermi operators and found that a 27~TeV machine will be able to probe scales of 
$\Lambda \gsim 8-15$~TeV and $\Lambda \gsim 4-5$~TeV for the scalar, vector and tensor $tu \ell \ell$ and $tc \ell \ell$ operators, respectively. Likewise, a 100 TeV proton collider will 
be sensitive to scales of $\Lambda \gsim 20-35$~TeV and $\Lambda \gsim 9-15$~TeV for the $tu \ell \ell$ and $tc \ell \ell$ operators.   

We furthermore explored potential searches for lepton non-universal effects that can be performed with our multi-lepton signals, finding e.g., that if the typical scale of these 4-Fermi 
$tu_i \ell \ell$ operators is around 5~TeV, then a separation of more than ${\cal O}(0.5)$~TeV between the scales of the $tu_i \mu\mu$ and $tu_i ee$ 4-Fermi terms, may be distinguishable via our di- and tri-lepton signatures, indicating that the underlying heavy physics is lepton non-universal. 

Finally, we end with a cautionary remark. A positive signal through these tests does not necessarily mean that the underlying new physics is flavor changing, but rather, it means that it may be so and further studies will be needed for confirmation.
\acknowledgments
The work of AS  was supported in part by the U.S. DOE contract \#DE-SC0012704.

\bibliographystyle{hunsrt.bst}
\bibliography{mybib2}

\end{document}